\documentclass[twocolumn]{aastex62}

\graphicspath{{./}{figures/}}

\shorttitle{Physical Conditions and Kinematics of OMC1}
\shortauthors{Teng \& Hirano}

\usepackage{amsmath}
\usepackage{hyperref}

\begin{document}

\title{Physical Conditions and Kinematics of the Filamentary Structure in Orion Molecular Cloud 1}

\correspondingauthor{Yu-Hsuan Teng}
\email{yuteng@ucsd.edu}

% \correspondingauthor{Naomi Hirano}
% \email{hirano@asiaa.sinica.edu.tw}

\author[0000-0003-4209-1599]{Yu-Hsuan Teng}
\affiliation{Institute of Astronomy and Astrophysics, Academia Sinica, \\
11F of Astronomy-Mathematics Building, National Taiwan University, No.1, Sec.4, Roosevelt Rd., Taipei 10617, Taiwan}
\affiliation{Department of Physics, National Taiwan University, \\
No.1, Sec.4, Roosevelt Rd., Taipei 10617, Taiwan}
\affiliation{Center for Astrophysics and Space Sciences, Department of Physics, University of California San Diego, \\
9500 Gilman Drive, La Jolla, CA 92093, USA}

\author[0000-0001-9304-7884]{Naomi Hirano}
\affiliation{Institute of Astronomy and Astrophysics, Academia Sinica, \\
11F of Astronomy-Mathematics Building, National Taiwan University, No.1, Sec.4, Roosevelt Rd., Taipei 10617, Taiwan}

\begin{abstract}

We have studied the structure and kinematics of the dense molecular gas in the Orion Molecular Cloud 1 (OMC1) region with the $\rm{N_2H^+}$ 3--2 line. The $6' \times 9'$ ($\sim 0.7 \times 1.1$ pc) region surrounding the Orion KL core has been mapped with the Submillimeter Array (SMA) and the Submillimeter Telescope (SMT). The combined SMA and SMT image having a resolution of $\sim 5.4''$ ($\sim 2300$ au) reveals multiple filaments with a typical width of $0.02$--$0.03$ pc.
On the basis of the non-LTE analysis using the $\rm{N_2H^+}$ 3--2 and 1--0 data, the density and temperature of the filaments are estimated to be $\sim 10^7$ $\rm{cm^{-3}}$ and $\sim 15$--$20$ K, respectively.
The core fragmentation is observed in three massive filaments, one of which shows the oscillations in the velocity and intensity that could be the signature of core-forming gas motions.
The gas kinetic temperature is significantly enhanced in the eastern part of OMC1, likely due to the external heating from the high mass stars in M42 and M43. 
In addition, the filaments are colder than their surrounding regions, suggesting the shielding from the external heating due to the dense gas in the filaments.
The OMC1 region consists of three sub-regions, i.e. north, west, and south of Orion KL, having different radial velocities with sharp velocity transitions.
There is a north-to-south velocity gradient from the western to the southern regions.
The observed velocity pattern suggests that dense gas in OMC1 is collapsing globally toward the high-mass star-forming region, Orion Nebula Cluster.
 
\end{abstract}


\section{Introduction} \label{sec:intro}

Filamentary structure has been commonly observed in star-forming clouds from parsec scale to sub-parsec scale \citep[e.g.][]{1979ApJS...41...87S, 2014prpl.conf...27A}. The prevalence of filamentary structure indicates its persistence for a large fraction of the lifetime of a star-forming cloud. Therefore, it is believed that such structure plays an important role in star formation process, and provides clues about the evolution of star-forming clouds. In addition, by examining the low-mass star-forming clouds within 300 pc, \citet{hub_filament} found that all young stellar groups are associated with ``hub-filament structure'', where the ``hub'' is a high column density region harboring young stellar groups, and the ``filaments'' are elongated structures with lower column density radiating from the hub. Such structure also exists in some distant regions that form high-mass stars, although its incidence in massive star-forming regions is still unclear \citep{hub_filament}. 

The Orion A molecular cloud, a large-scale filament with an integral-like shape, is the nearest high-mass star-forming region at a distance of 414 pc \citep{omc1_dist}. The Orion Molecular Cloud 1 (OMC1), residing at the center of the Orion A, is the most massive component ($>2200 M_\odot$) and the most active star-forming region in the Orion Molecular Cloud \citep{omc_fil}. Previous VLA observations in $\rm{NH_3}$ \citep{omc1_nh3} revealed a typical hub-filament structure in OMC1, in which several filaments radiate from the Orion KL. These filaments appear to be hierarchical: the large-scale filament is consist of narrower filaments in small scale. Recent high-resolution observation with ALMA in $\rm{N_2H^+}$ J=1--0 \citep{omc1_alma} resolved a total of 28 filaments with a FWHM of $\sim 0.02$--$0.05$ pc in OMC1, and the cores inside these small-scale filaments are possible sites for star formation. Therefore, studying the physical conditions and gas motions in OMC1 is likely the key to understand the evolution of hub-filament structure and its relation with star formation. 

We present our observations toward the OMC1 region in $\rm N_2H^+$ 3--2 with an angular resolution of $\sim 5.4''$ obtained using the Submillimeter Array (SMA) and the Submillimeter Telescope (SMT). Further analyses using the $\rm N_2H^+$ 1--0 data provided by \citet{omc1_alma} are also presented. The rotational transitions of $\rm N_2H^+$ are known to be good tracers of dense and quiescent gas. With a critical density of $\sim 10^6$ $\rm cm^{−3}$, which is higher than that of ammonia studied by \citet{omc1_nh3}, $\rm N_2H^+$ 3--2 can probe the dense gas inside the sub-parsec scale filaments that are directly related to star formation.
In addition, $\rm N_2H^+$ is less affected by depletion even in the dense and cold environments, and is also less affected by dynamic processes such as outflows and expanding H{\small \ II} regions.

This study investigates the structure, physical conditions, and gas motions in the OMC1 region. We describe the details of our observations and data reduction in Section \ref{sec:obs}, and present the results in Section \ref{sec:result}. The structural properties, physical conditions and gas kinematics are analyzed in Section \ref{sec:analysis}. Finally, we discuss the implications of these results in Section \ref{sec:discussion}, and summarize our conclusions in Section \ref{sec:conclusion}.

\section{Observations and Data Reduction} \label{sec:obs}

\subsection{1.1 mm Observations with the SMA}

\begin{figure} %[ht!]
\plotone{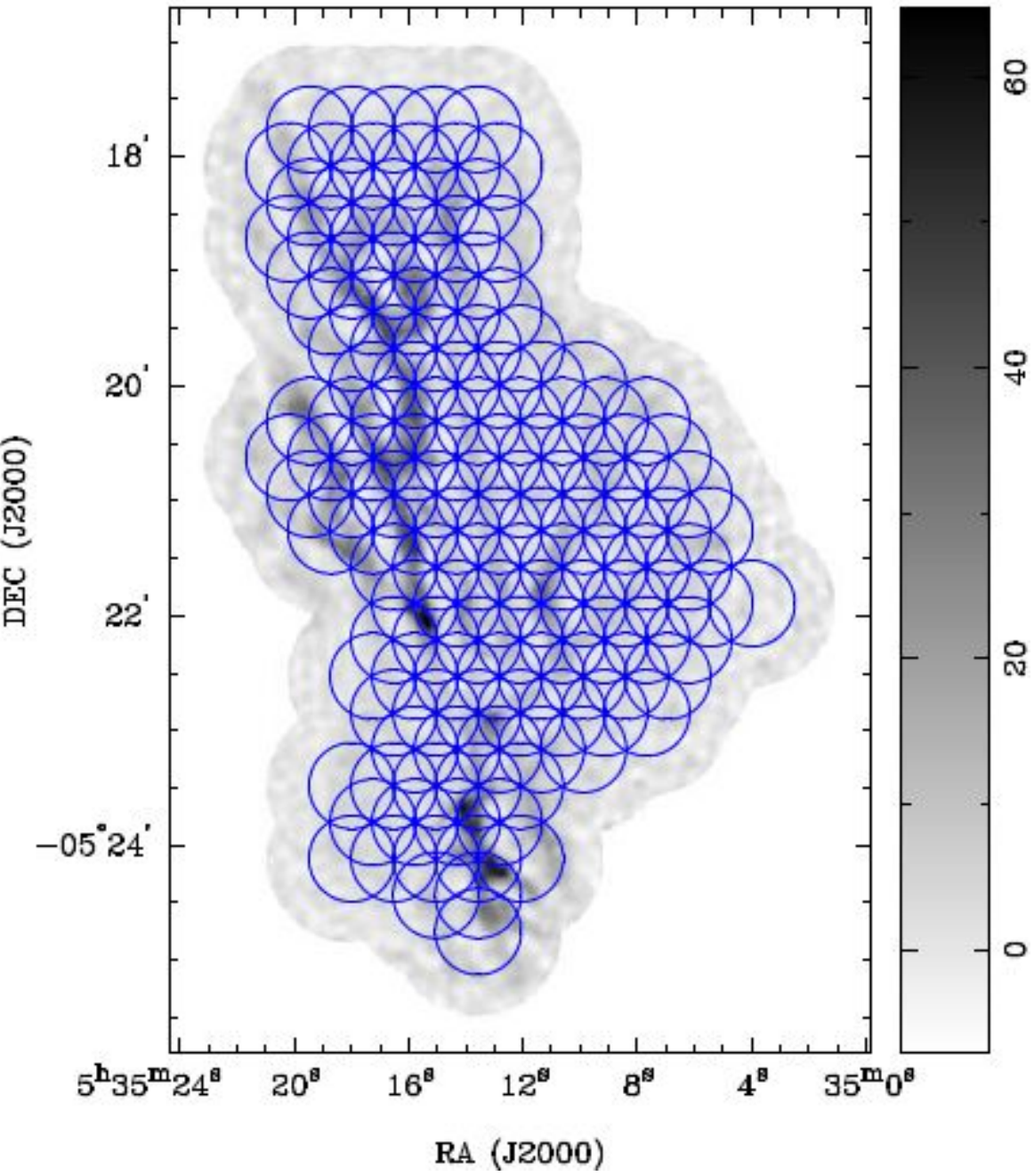}
\caption{Integrated intensity map made by only the SMA observation. The 144 pointings covering the OMC1 region are overlaid on the SMA map.
\label{fig:pointing}}
\end{figure}

Observations of the $\rm N_2H^+$ J=3--2, $\rm HCO^+$ J=3--2, and HCN J=3--2 lines together with the 1.1 mm continuum were carried out with the SMA on February 14, 20, and 24, 2014. A sub-compact configuration with six antennas in the array was used, providing baselines ranging from 9.476 m to 25.295 m. The shortest and longest uv distances are 5.6 $\rm k\lambda$ and 23.6 $\rm k\lambda$, repectively. The primary beam of the 6-m antennas has a size of $42''$ (HPBW), and the synthesized beam size is $5.53'' \times 5.25''$. The bandwidth was $4$ GHz per sideband, and the frequency resolution was 203 kHz that corresponds to the velocity resolution of $\sim 0.22$ $\rm km\ s^{-1}$ at the rest frequency of $\rm N_2H^+$ 3--2. Using 144 pointing mosaic with a Nyquist sampled hexagonal pattern, as shown in Figure \ref{fig:pointing}, the observed area covered $\sim 5' \times 7'$.
In order to obtain uniform $uv$-coverage for 144 pointngs, we observed each pointing for 5 seconds and visited all pointings in a loop.
Each of the 144 pointings was visited three times in each observing run, giving a total on-source integration time per pointing of 45 seconds.

The visibility data were calibrated using the MIR/IDL software package.\footnote{\url{https://www.cfa.harvard.edu/rtdc/SMAdata/process/mir/}} The gain calibrators were 0501-019 and 0607-085, the flux calibrator was Ganymede on Feb. 14 and Callisto on other two days, and the bandpass calibrator was 3C279. 
The image processing was carried out using the MIRIAD package \citep{miriad}. The image cube was generated with Briggs weighting with a robust parameter of 0.5, followed by a nonlinear joint deconvolution using the CLEAN-based algorithm, MOSSDI.
The final data cube has a rms noise level of $\sim 0.5$ K for a $\sim 0.22$ $\rm km\ s^{-1}$ velocity channel.
In this paper, we focus on the results and analyses of $\rm N_2H^+$ line. Results of $\rm HCO^+$ and HCN will be presented in a forthcoming paper.

\subsection{1.1 mm Observations with the SMT}

We simultaneously observed the $\rm N_2H^+$ J=3--2 and $\rm HCO^+$ J=3--2 lines on November 16 and 17, 2018 using the SMT of the Arizona Radio Observatory. We used the SMT 1.3 mm ALMA band 6 receiver and the filter-bank backend. The beam size is $28.45''$ in HPBW at the frequency of $\rm N_2H^+$ 3--2, and the main-beam efficiency is $0.71 \pm 0.05$. The On The Fly (OTF) mode was used in order to cover the mapping area of $6' \times 9'$ centered at R.A.(J2000) = $\rm 5^h 35^m 12^s.1$ and Decl.(J2000) = $-5^\circ 21' 15''.4$. The data were reduced with CLASS.\footnote{\url{http://www.iram.fr/IRAMFR/GILDAS}} The rms noise levels are $\sim 0.3$ K at a spectral resolution of 250 kHz ($\sim 0.27$ $\rm km\ s^{-1}$).

\subsection{SMA and SMT Data Combination}

We combined the data cubes of the SMA and SMT by using the MIRIAD task \textit{immerge}. 
The method of this task, known as feathering, merges linearly two images with different resolutions in their Fourier domain (i.e. spatial frequency).
In the case of combining the single-dish and mosaicing data, \textit{immerge} gives unit weight to the single-dish data at all spatial frequencies, and tapers the low spatial frequencies of the mosaicing data, so as to produce the gaussian beam of the combined data equal to that of the mosaicing data.
Inputs of \textit{immerge} include the ``CLEANed'' SMA image cube, SMT image cube, and the flux calibration factor. 
After checking the consistency in the flux scale of two input image cubes, we set the flux calibration factor to 1. 
The integrated intensity (moment 0) map and the intensity-weighted radial velocity (moment 1) map after combination are shown in Figure \ref{fig:obs_combine}.
The angular resolution and the rms noise level of the combined map are $\sim 5.4''$ and $1.0$ $\rm K \cdot km\ s^{-1}$, respectively.

% \subsection{$N_2H^+$ J=1--0 Observations with NRO 45-m}

% In this paper, we also use the public $\rm N_2H^+$ 1--0 data observed with the 45-m telescope of the Nobeyama Radio Observatory (NRO).\footnote{\url{https://www.nro.nao.ac.jp/~kt/html/fits.html}} The HPBW and the main beam efficiency of the 45-m telescope were $17.8''$ and $0.51$, respectively, at $93$ GHz. The spectral resolution was $37.8$ kHz, the corresponding velocity resolution was $\sim 0.12$ $\rm km\ s^{-1}$. Further observational details can be found in \citet{omc1_nro45}. 

\subsection{$N_2H^+$ J=1--0 Observations with ALMA and IRAM 30-m}

To analyze the physical properties of the filamentary structure in high resolution, we use the $\rm N_2H^+$ 1--0 data cube provided by \citet{omc1_alma}, where the ALMA and IRAM 30-m data were combined.\footnote{\url{https://doi.org/10.7910/DVN/DBZUOP}} The ALMA + IRAM 30-m combined image cube has a circular beam with a size of $4.5''$ in FWHM, and a rms level of $25$ $\rm mJy\ beam^{-1}$ at a spectral resolution of $0.1$ $\rm km\ s^{-1}$. The observational details are described in \citet{hacar_2017,omc1_alma}.

\section{Results} \label{sec:result}

% \begin{figure} %[ht!]
% %\plotone{ra2vel.pdf}
% \includegraphics[width=\linewidth]{ra2vel.pdf}

\begin{figure*}
\begin{minipage}{.48\linewidth}
\centering
\includegraphics[width=\linewidth]{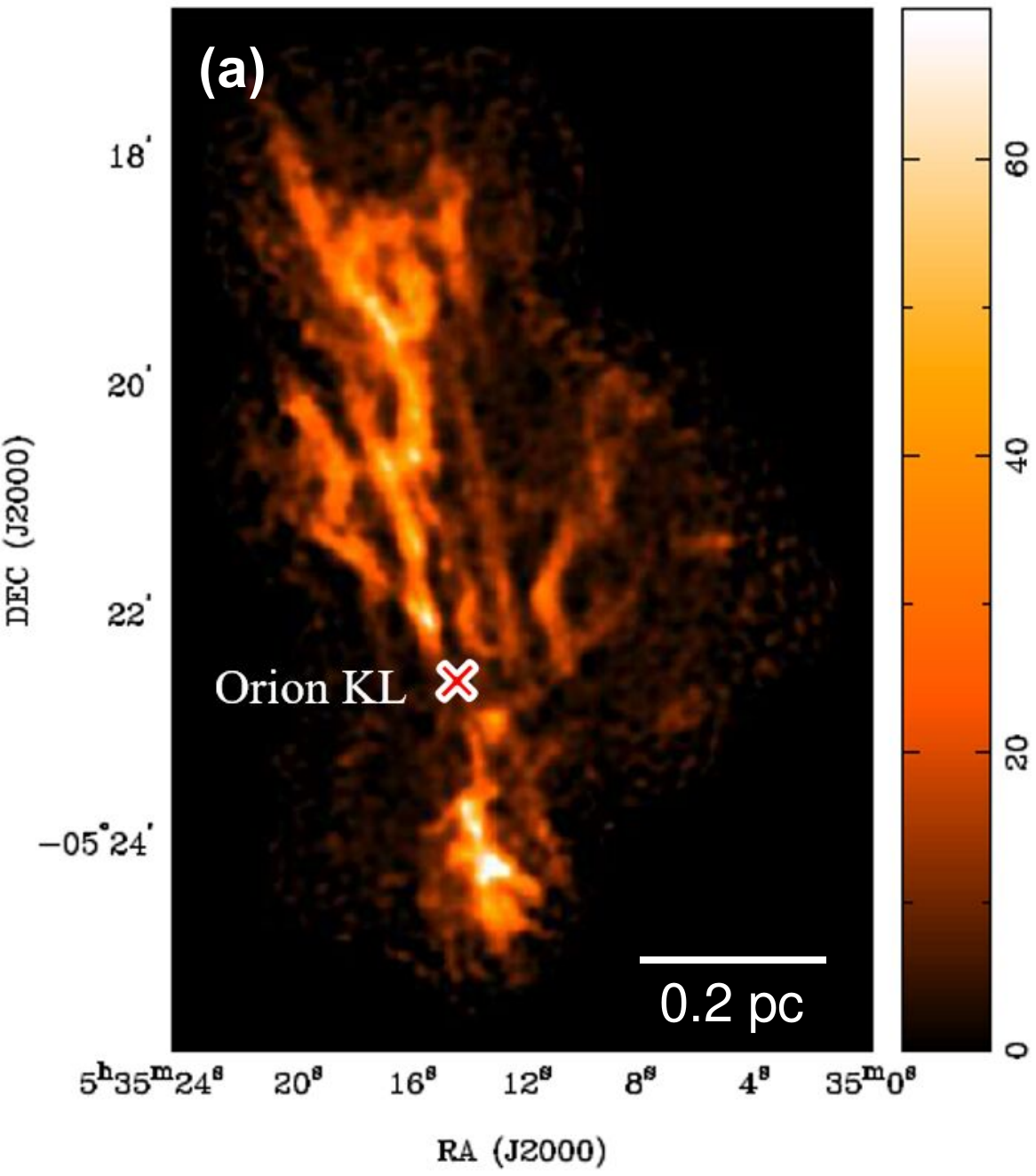}%\\
%(a)
\end{minipage}
\hfill
\begin{minipage}{.48\linewidth}
\centering
\includegraphics[width=\linewidth]{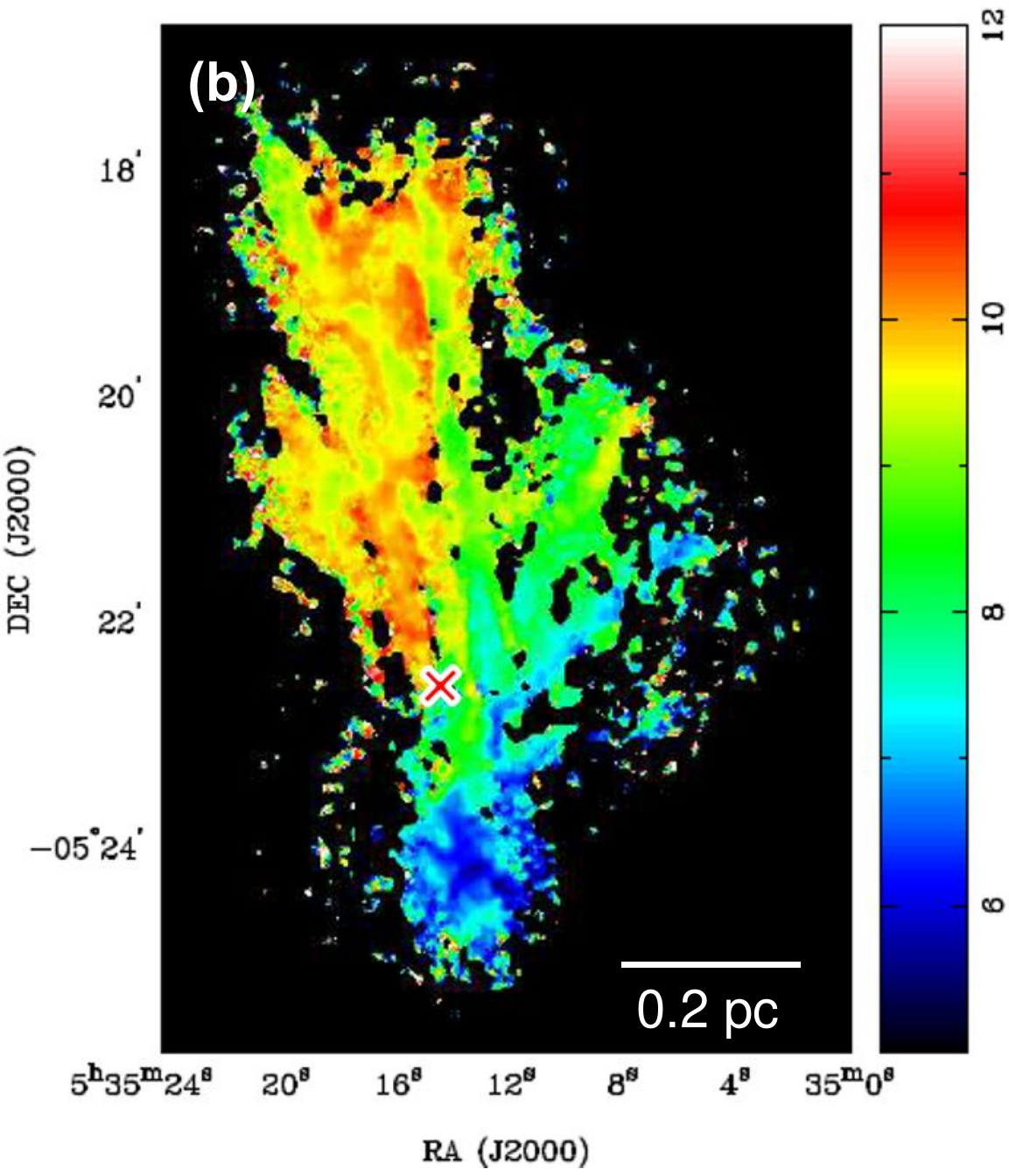}%\\
%(b)
\end{minipage}
\caption{Moment 0 ($\rm K \cdot km\ s^{-1}$) and moment 1 ($\rm km\ s^{-1}$) of the combined SMA and SMT data in $\rm N_2H^+$ 3--2. The cross depicts the position of Orion KL at (R.A., Decl.) = ($\rm 5^h 35^m 14^s.5$, $-5^\circ 22' 30''$).}
\label{fig:obs_combine}
\end{figure*}

Figure \ref{fig:obs_combine} presents the moment 0 (integrated intensity) and moment 1 (intensity-weighted radial velocity) maps of the SMA + SMT combined image. 
Figure \ref{fig:obs_combine}a shows that most of the emission comes from the filamentary structure having a typical FWHM of $0.02$--$0.03$ pc. Several high-intensity and clumpy structures can also be seen inside the filaments. 
Orion KL at R.A.(J2000) = $\rm 5^h 35^m 14^s.5$ and Decl.(J2000) = $-5^\circ 22' 30''$ is near the center of the maps. Different from observations in continuum or most other molecular lines, there is no significant $\rm{N_2H^+}$ emission from the Orion KL region due to the destruction of $\rm N_2H^+$ molecules in active regions. 

Figure \ref{fig:obs_combine}b reveals that the radial velocity distribution shows a trimodal pattern; the bright filaments to the north of Orion KL have a velocity range of $\sim$9--11 km s$^{-1}$ (hereafter, referred to as the northern region), the fainter filaments extending to the northwest are seen at $\sim$7--9 km s$^{-1}$ (western region), and the ones to the south of Orion KL are at $\sim$5--7 km s$^{-1}$ (southern region, also known as OMC1-South). These three regions with different velocities converge at the Orion KL region.
The velocity difference between the northern and western region was also reported by \citet{omc1_nh3} and \citet{Monsch_2018} based on their NH$_3$ observations. A clearer analysis for the velocity transition among the three regions will be presented in Section \ref{subsec::fitting}.

\begin{figure} %[ht!]
%\plotone{ratio_mom0_overlap_bar.pdf}
\includegraphics[width=\linewidth]{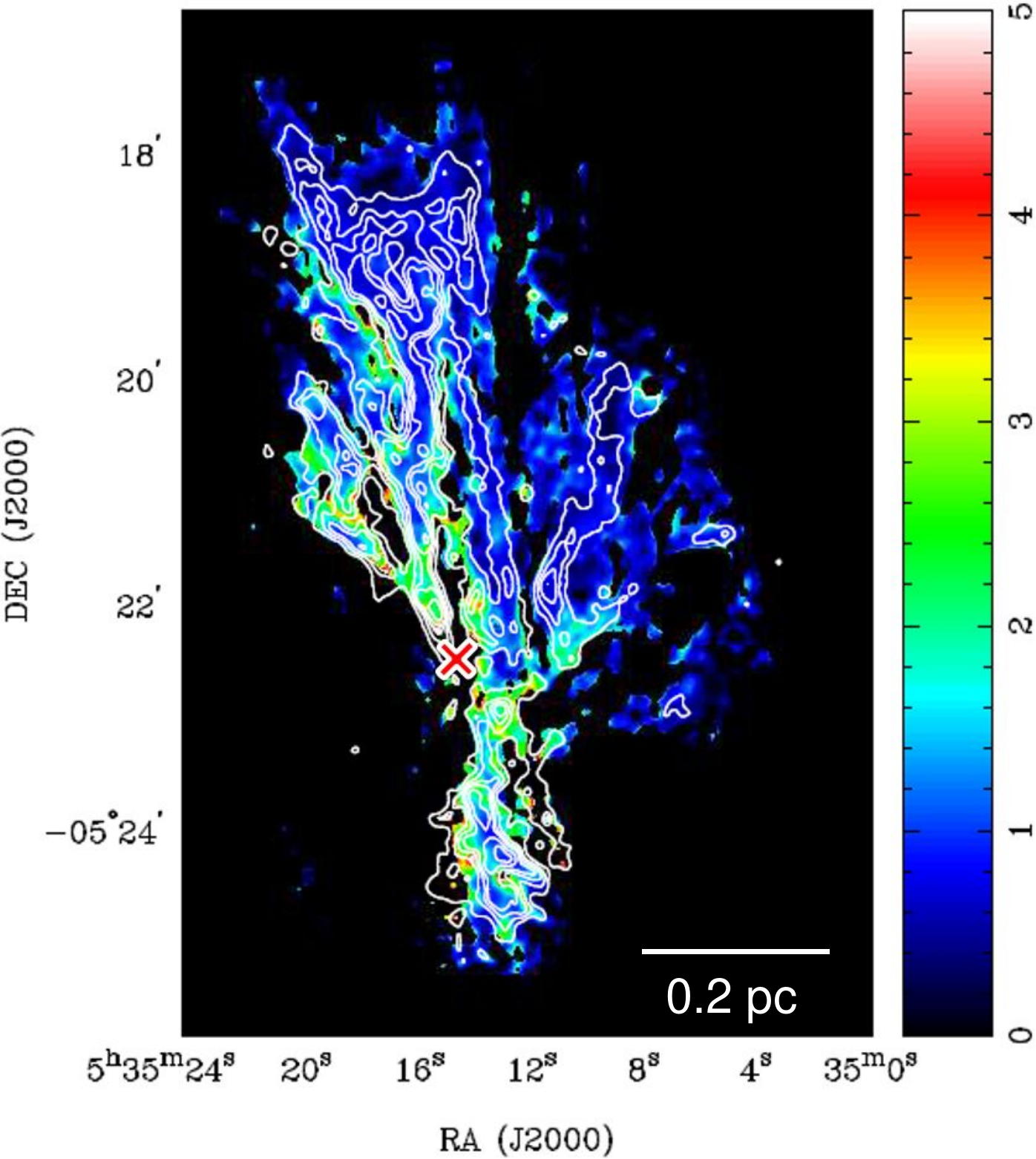}
\caption{$\rm N_2H^+$ 3--2/1--0 intensity ratio map using 1--0 image observed with ALMA + IRAM 30-m \citep{omc1_alma}. Contour levels of $\rm N_2H^+$ 3--2 moment 0 are overlaid on the line ratio map.}
\label{fig:ratio}
\end{figure}

In order to derive physical conditions in different sub-regions, we use the ALMA + IRAM 30-m image in $\rm N_2H^+$ 1--0 for analyses in Section \ref{subsec::non-lte}.
The high-resolution 3--2/1--0 ratio map (Figure \ref{fig:ratio}) was made using the SMA + SMT image and the ALMA + IRAM 30-m image convolved to the same beam size as the SMA + SMT image.
Figure \ref{fig:ratio} reveals that the 3--2/1--0 ratio is overall higher in the eastern part of OMC1. In addition, the 3--2/1--0 ratio tends to be lower in the filament regions as compared to the surrounding non-filament regions. 
The typical line ratio in the filament regions is $\sim 1.0$ even in the eastern part of OMC1, while that of the non-filament region is $\sim 2.2$.
As different ratios may imply different physical conditions, we determine the physical parameters of the filament and non-filament regions in Section \ref{subsec::non-lte}.

\section{Analysis} \label{sec:analysis}

\subsection{Structural Properties of OMC1} \label{sec::structure}

\begin{figure*}
\begin{minipage}{.48\linewidth}
\centering
\includegraphics[width=\linewidth]{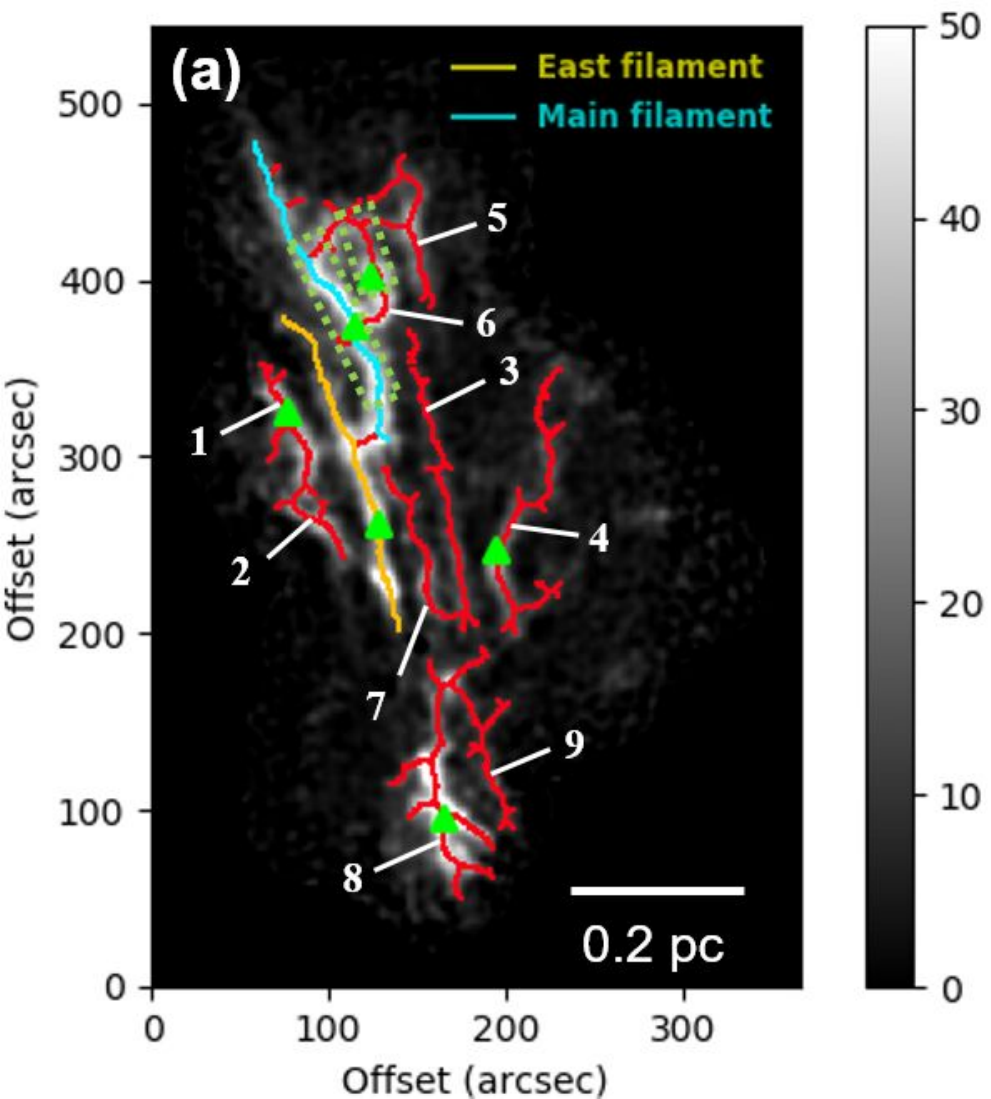}%\\
%(a)
\end{minipage}
\hfill
\begin{minipage}{.48\linewidth}
\centering
\includegraphics[width=\linewidth]{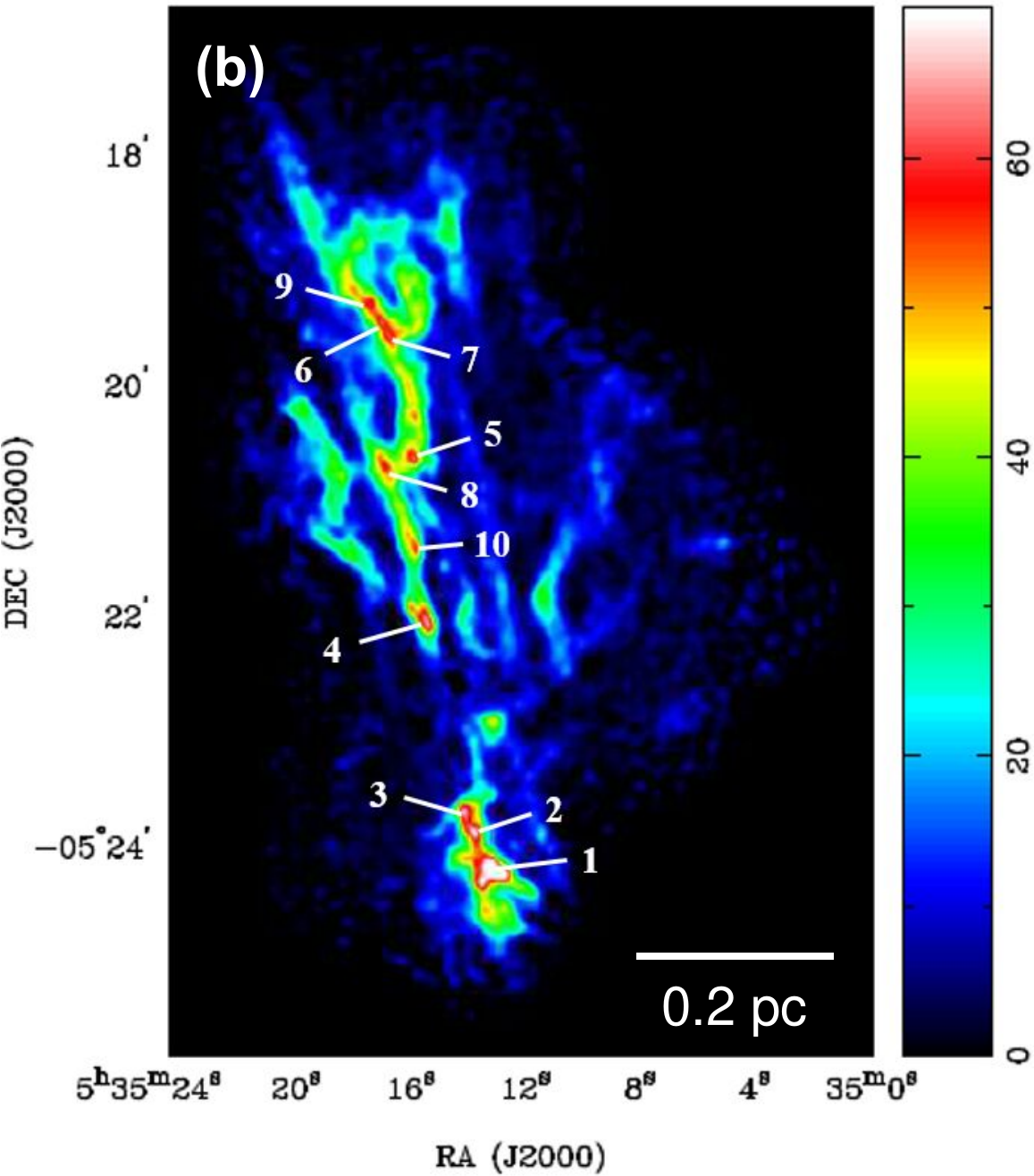}%\\
%(b)
\end{minipage}
\caption{(a) Filaments identified using the \textit{FilFinder}, including the main (blue) and east (yellow) filaments and others (red). The dashed green boxes enclose the regions where multiple velocity components are presented, and the green triangles mark the locations where spectra are shown in Figure \ref{fig:spec}.}  (b) Positions of the 10 cores identified using \textit{2-D Clumpfind}.
\label{fig:structure}
\end{figure*}

\subsubsection{Filament Identification}
Filament identification is done by applying the python package \textit{FilFinder} \citep{filfinder} to the SMA + SMT combined moment 0 map. The \textit{FilFinder} algorithm segments filamentary structure by using adaptive thresholding, which performs thresholding over local neighborhoods and allows for the extraction of structure over a large dynamic range. 
Input parameters for \textit{FilFinder} include (1) global threshold---minimum intensity for a pixel to be included; (2) adaptive threshold---the expected full width of filaments for adaptive thresholding; (3) smooth size---scale size for removing small noise variations; (4) size threshold---minimum number of pixels for a region to be considered as a filament. 

We set 9 $\rm K \cdot km\ s^{-1}$ as the global threshold. To focus on filaments with length $\gtrsim 0.1$ pc, we set the size threshold as 400 square pixels, where the pixel size is 1 $\rm arcsec^2$. The adaptive threshold is set to 0.06 pc, which is approximately twice the width of the filaments, and the smooth size is set to 0.03 pc.
We find the result matches better with identification by human eyes when the smooth size is set to $\sim 0.5$ times the adaptive threshold. The smaller we set the smooth size, the more short branches would be identified. On the other hand, a larger smooth size tends to make filaments connected since a larger regions of data are smoothed.

In total, 11 filaments have been identified.
Figure \ref{fig:structure}a shows the result of filament identification, where the gray-scale image is the combined moment 0 map, and the colored lines are the identified major axes for each filament. 
The identified filaments and their lengths and widths are listed in Table \ref{tab:fil_ident}, together with the physical properties described in Section \ref{subsec:core_formation}.
Since the hyperfine components in $\rm N_2H^+$ 3--2 cannot be separated, the identification is based on the moment 0 map instead of 3-D datacube. 
We have estimated the typical FWHM of the identified filaments by fitting Gaussian to the several cuts perpendicular to the filaments. The FWHM of these filaments are $0.02$--$0.03$ pc, which is consistent with those identified in \citet{omc1_alma} based on the ALMA + IRAM 30-m observations in $\rm N_2H^+$ 1--0.

\citet{omc1_alma} has identified the filaments in Orion A with a different algorithm based on the 3-D datacube. 
Since we identified the filaments in 2-D using the moment 0 map, the results are different from those of 3-D if there are multiple velocity components in the same line of sight.
For example, our main filament and filament 6, which can be clearly seen in the moment 0 map are not identified by \citet{omc1_alma}. As shown in Figure \ref{fig:spec}a and \ref{fig:spec}b, these two filaments contain multiple velocity components along the line of sight, which could cause a difference between results of 2D- and 3D-based algorithms.
Apart from these filaments, other OMC1 filaments show single velocity component in the spectra.
Therefore, there is no significant difference in the results between 2-D and 3-D identifications. 

%Among the filaments in OMC1, 82\% of them consist of a single velocity component, and 18\% have multiple components. 

The moment 0 images in $\rm N_2H^+$ reveal three filaments with high-intensity clumpy cores, one of which is in the OMC1-South filament 8. 
We will refer to the two prominent filaments in the northern region as the \textit{main filament} and the \textit{east filament}, which are shown in blue and yellow, respectively, in Figure \ref{fig:structure}a. The gas kinematics inside these filaments will be analyzed in Section \ref{subsec:kinematics}. 

\begin{table*}
\centering
\caption{Properties of the identified filaments \label{tab:fil_ident}}
\begin{tabular}{lccccccc}
  \hline\hline
  Filament &Length &$\rm Width_{FWHM}$ &$\rm M_{lin}$ &$\Delta v$ &$\Delta v_{\rm nt}/c_s(T)$ &$\rm M_{crit}$ &$\rm M_{lin}/M_{crit}$\\
  &($\pm 0.001$pc) &($\pm 0.001$pc) &($M_\odot\ pc^{-1}$) &($\rm km\ s^{-1}$) & &($M_\odot\ pc^{-1}$) & \\
  \hline
  Main &0.349 &$0.025$ &94.2--101.7 &$1.0\pm0.2$ &1.5 &119.7 &0.79--0.85\\
  East &0.365 &$0.023$ &78.5--85.6 &$0.9\pm0.3$ &1.3 &103.8 &0.76--0.83\\
  1 &0.159 &$0.024$ &84.0 &$0.6\pm0.2$ &0.85 &66.0 &1.27\\
  2 &0.100 &$0.023$ &76.2 &$0.7\pm0.3$ &1.0 &76.9 &0.99\\
  3 &0.279 &$0.017$ &42.5 &$1.3\pm0.6$ &1.9 &177.6 &0.24\\
  4 &0.303 &$0.020$ &61.8 &$1.1\pm0.3$ &1.6 &137.3 &0.45\\
  5 &0.155 &$0.024$ &84.0 &$1.0\pm0.5$ &1.5 &119.7 &0.70\\
  6 &0.124 &$0.033$ &166.3 &$1.1\pm0.3$ &1.6 &137.3 &1.21\\
  7 &0.175 &$0.022$ &68.9 &$1.0\pm0.4$ &1.5 &119.7 &0.58\\
  8 &0.237 &$0.022$ &81.5--121.0 &$2.0\pm0.9$ &2.95 &371.6 &0.22--0.33\\
  9 &0.159 &$0.017$ &42.5 &$1.0\pm0.3$ &1.5 &119.7 &0.36\\
  \hline
\end{tabular}
\end{table*}

\subsubsection{Core Identification} \label{subsubsec::core_ident}

We identify the high-intensity cores inside the filaments by using the 2-D version of \textit{Clumpfind} \citep{clfind}. The \textit{Clumpfind} algorithm contours input data with the values assigned by users, and then distinguishes each core region by those values. It starts from the highest contour level, and then works down through the lower levels, finding new cores and extending previously defined ones until the lowest contour level is reached. 
We use the combined SMA + SMT integrated intensity map as the input data, and set the contour levels as 50, 53.6, 57.2, 60.8, 64.4 $\rm K \cdot km\ s^{-1}$. The spacing of the contour levels are set constantly as $\Delta T = 2 T_{\rm rms} = 3.6$, which could lower the the percentage of false detection to $< 2\%$ as suggested in \citet{clfind}. 

Figure \ref{fig:structure}b shows the positions of the 10 cores identified by \textit{2-D Clumpfind}. 
Table \ref{tab:core_ident} lists the properties of these cores including position, peak flux, and effective radius ($\rm R_{\rm eff}$), which are all direct outputs of \textit{2-D Clumpfind}. Note that the $R_{\rm eff}$ in \textit{Clumpfind} is defined as $R_{\rm eff} = \sqrt{A/\pi}$, where $A$ is the area with emission above the criteria. In Table \ref{tab:core_ident}, the linewidth ($\Delta v$) is determined by hyperfine spectral fitting (see Section \ref{subsec::fitting}), the mass ($\rm M_{core}$) is derived by using the densities determined from non-LTE analysis (see Section \ref{subsec::non-lte}), and the virial mass ($\rm M_{vir}$) is calculated from the linewidth using Equation \ref{def_Mvir} (see Section \ref{subsec:core_formation}).  

\begin{table*}
\centering
\caption{Properties of the identified cores \label{tab:core_ident}}
\begin{tabular}{lcccccccc}
  \hline\hline
  Core &R.A. &Decl. &Peak flux &$R_{\rm eff}$ &$\Delta v$ &$\rm M_{core}$ &$\rm M_{vir}$ \\
  &(J2000.0) &(J2000.0) &($\rm K \cdot km\ s^{-1}$) &(arcsec) &($\rm km\ s^{-1}$) &($M_\odot$) &($M_\odot$) \\
  \hline
  1 &05:35:13.10 &-5:24:12.4 &84.3 &8.9 &3.5 &14.4--45.6 &46.1 \\
  2 &05:35:13.70 &-5:23:52.4 &69.4 &5.6 &2.8 &3.6--11.3 &18.5 \\
  3 &05:35:14.10 &-5:23:41.4 &66.5 &4.7 &1.6 &2.0--6.4 &5.0 \\
  4 &05:35:15.37 &-5:22:04.4 &66.0 &5.1 &2.0 &2.6--8.2 &8.5 \\
  5 &05:35:15.84 &-5:20:38.4 &63.4 &3.5 &0.9 &0.9--2.8 &1.2 \\
  6 &05:35:16.90 &-5:19:27.4 &61.9 &4.0 &-- &1.3--4.2 &-- \\
  7 &05:35:16.70 &-5:19:34.4 &60.8 &4.4 &-- &1.7--5.3 &-- \\
  8 &05:35:16.84 &-5:20:42.4 &59.2 &3.6 &2.2 &0.9--3.0 &7.4 \\
  9 &05:35:17.37 &-5:19:17.4 &58.7 &4.2 &-- &1.5--4.7 &-- \\
  10 &05:35:15.77 &-5:21:24.4 &57.1 &2.6 &1.1 &0.4--1.1 &1.3 \\
  \hline
\end{tabular}
\end{table*}

\subsection{Hyperfine Spectral Fitting} \label{subsec::fitting}

To determine the physical parameters, we conduct hyperfine spectral fitting on the combined SMA and SMT data for the regions with $S/N > 5$. We fit the $V_{\rm LSR}$, linewidths ($\Delta v$), excitation temperatures ($T_{\rm ex}$), and total opacities ($\tau_{\rm tot}$) under the assumption of local thermodynamic equilibrium (LTE).
With the relative intensities among 16 main hyperfine components for the $\rm N_2H^+$ 3--2 line \citep{hfs_32}, the opacity of each component is assumed as a Gaussian profile 
\begin{eqnarray}
&&\tau_i(v) = \tau_i \exp \left[-4 \ln 2 \left( \frac{v - v_i - v_{\rm sys}}{\Delta v} \right)^2 \right]
\label{eqn_tau_i} 
\end{eqnarray}
where $v_i$ is the velocity offset from the reference component and $v_{\rm sys}$ is the systemic velocity.
Then, we obtain the optical depths of the multiplets as 
\begin{eqnarray}
&&\hspace{-0.3in}\tau(v) = \tau_{\rm tot} \sum_{i=1}^{16} R_i \exp \left[-4 \ln 2 \left( \frac{v - v_i - v_{\rm sys}}{\Delta v} \right)^2 \right]
\label{eqn_tau_v}
\end{eqnarray}
where $R_i$ is the relative intensity for the $i$th hyperfine component, and $\tau_i = \tau_{\rm tot} R_i$.
The brightness temperature at each pixel can be represented as 
\begin{eqnarray}
&&T_{b}(v) = [J(T_{\rm ex}) - J(T_{\rm bg})] [1 - \exp(-\tau(v))]
\label{Tb_lte}
\end{eqnarray}
where $T_{\rm bg}$ is the cosmic background temperature (2.73 K), and
\begin{eqnarray}
&&J(T) = \frac{\frac{h \nu}{k}}{\exp \left( \frac{h \nu}{k T} \right) - 1}
\label{def_J}
\end{eqnarray}

\begin{figure*}
\begin{tabular}{lccc}
\includegraphics[width=0.32\linewidth]{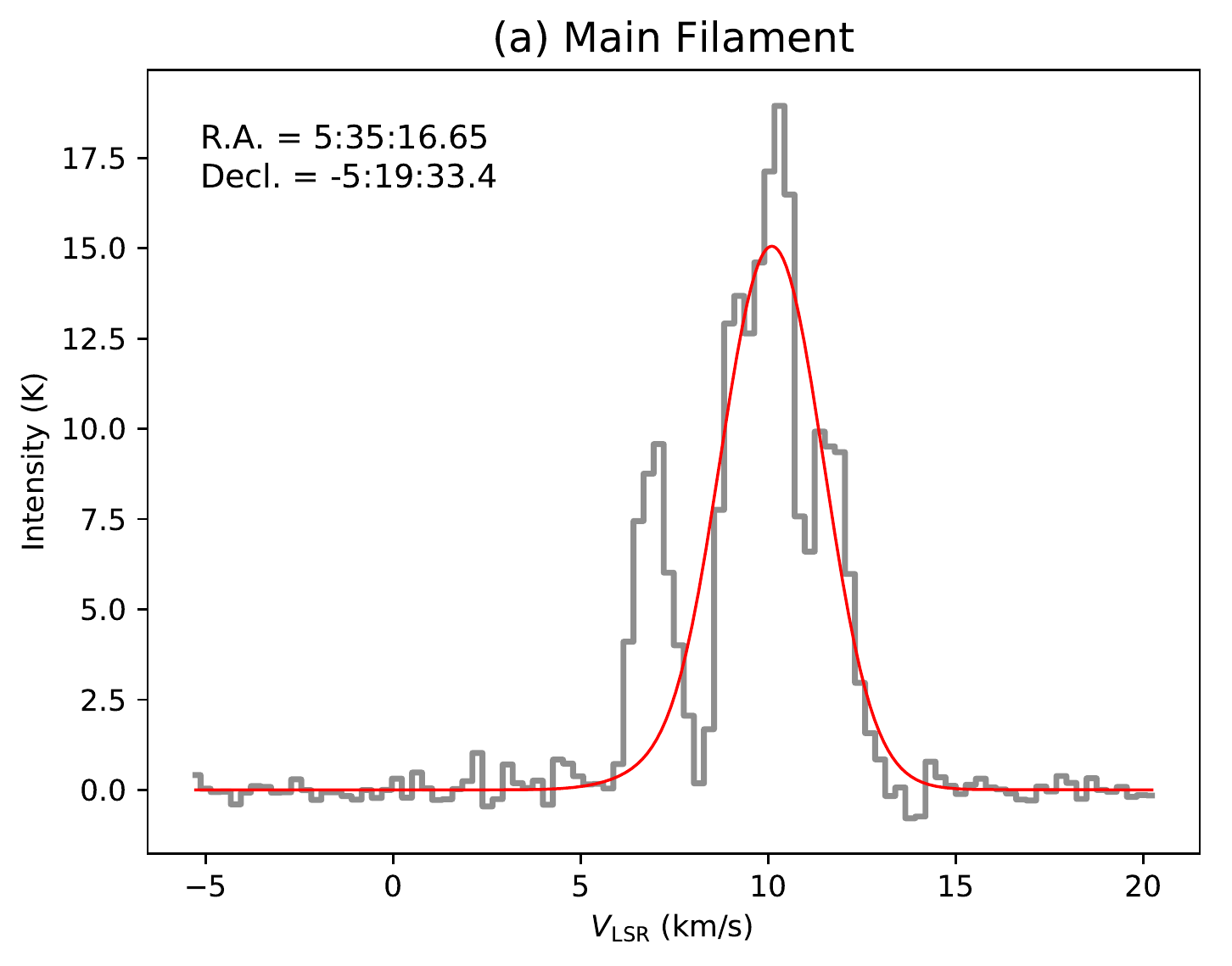} &\includegraphics[width=0.32\linewidth]{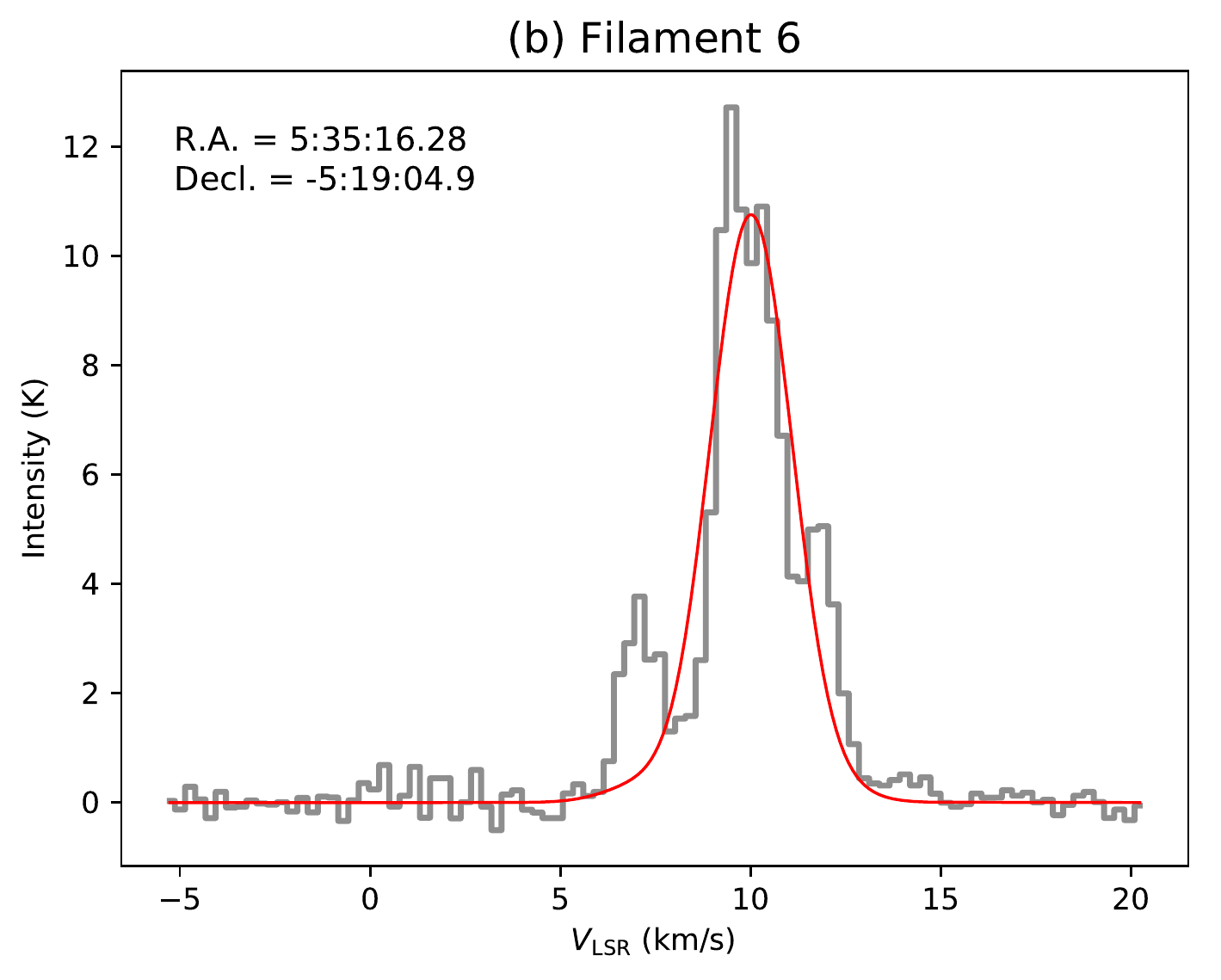} &\includegraphics[width=0.32\linewidth]{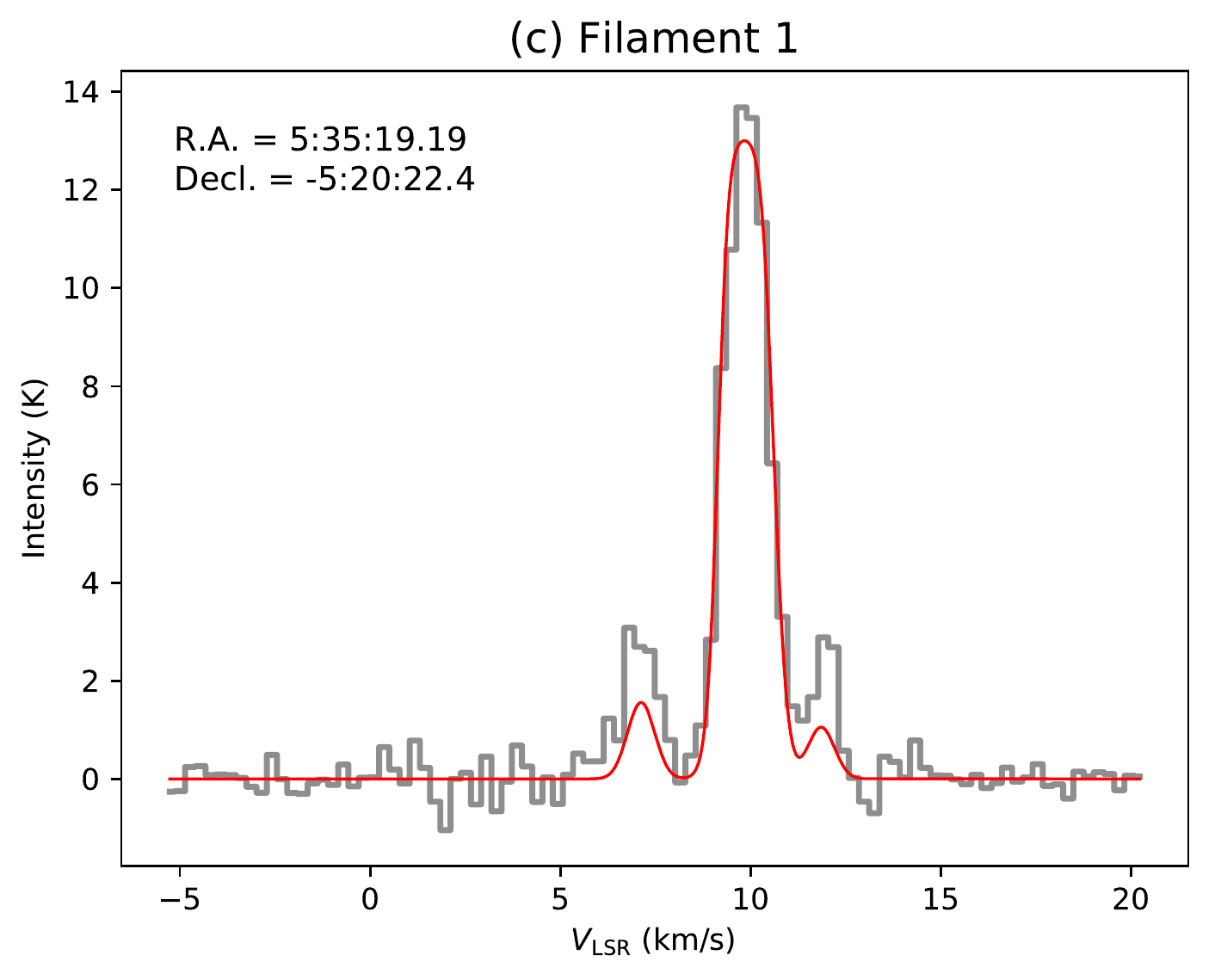} \\
\includegraphics[width=0.32\linewidth]{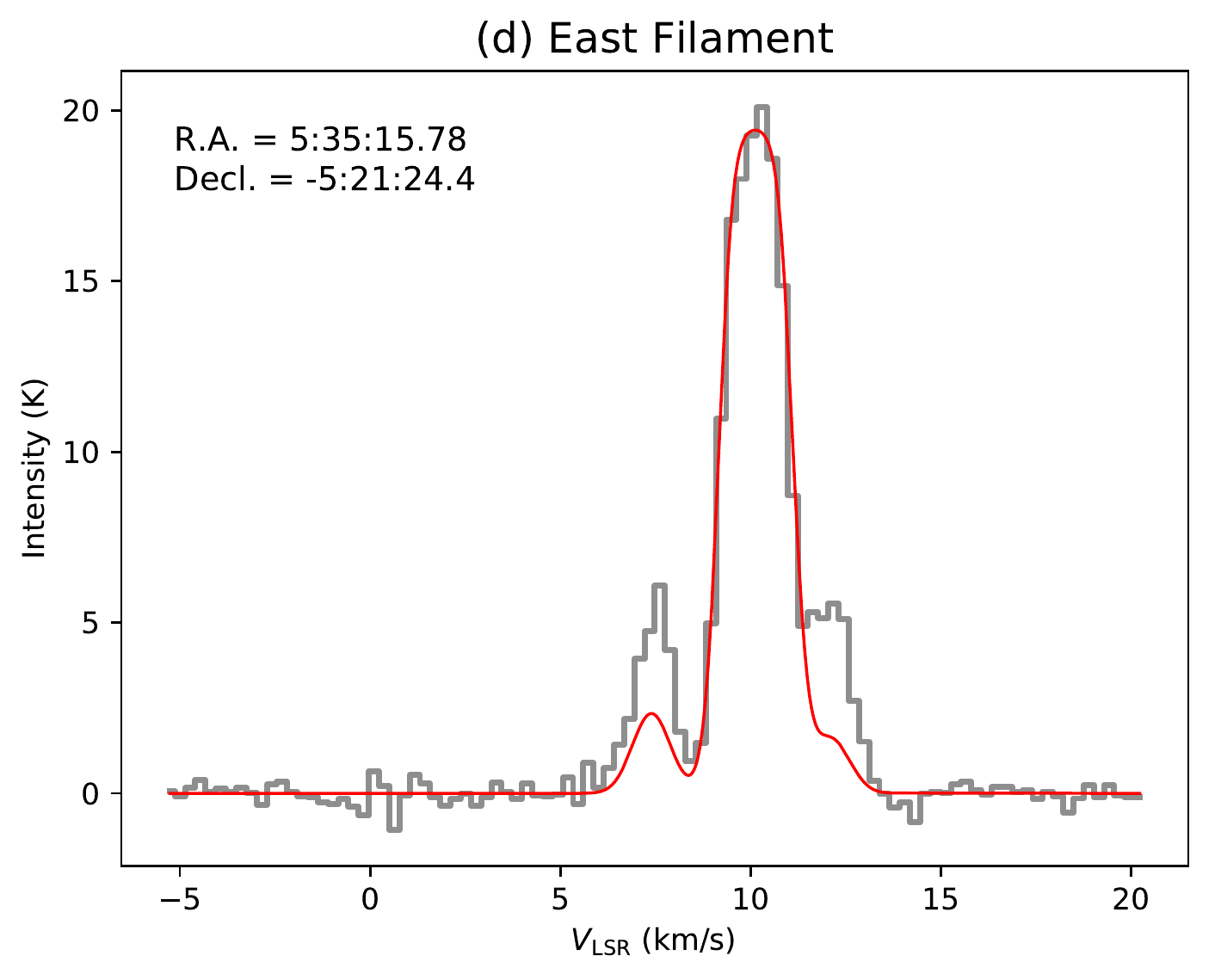} &\includegraphics[width=0.32\linewidth]{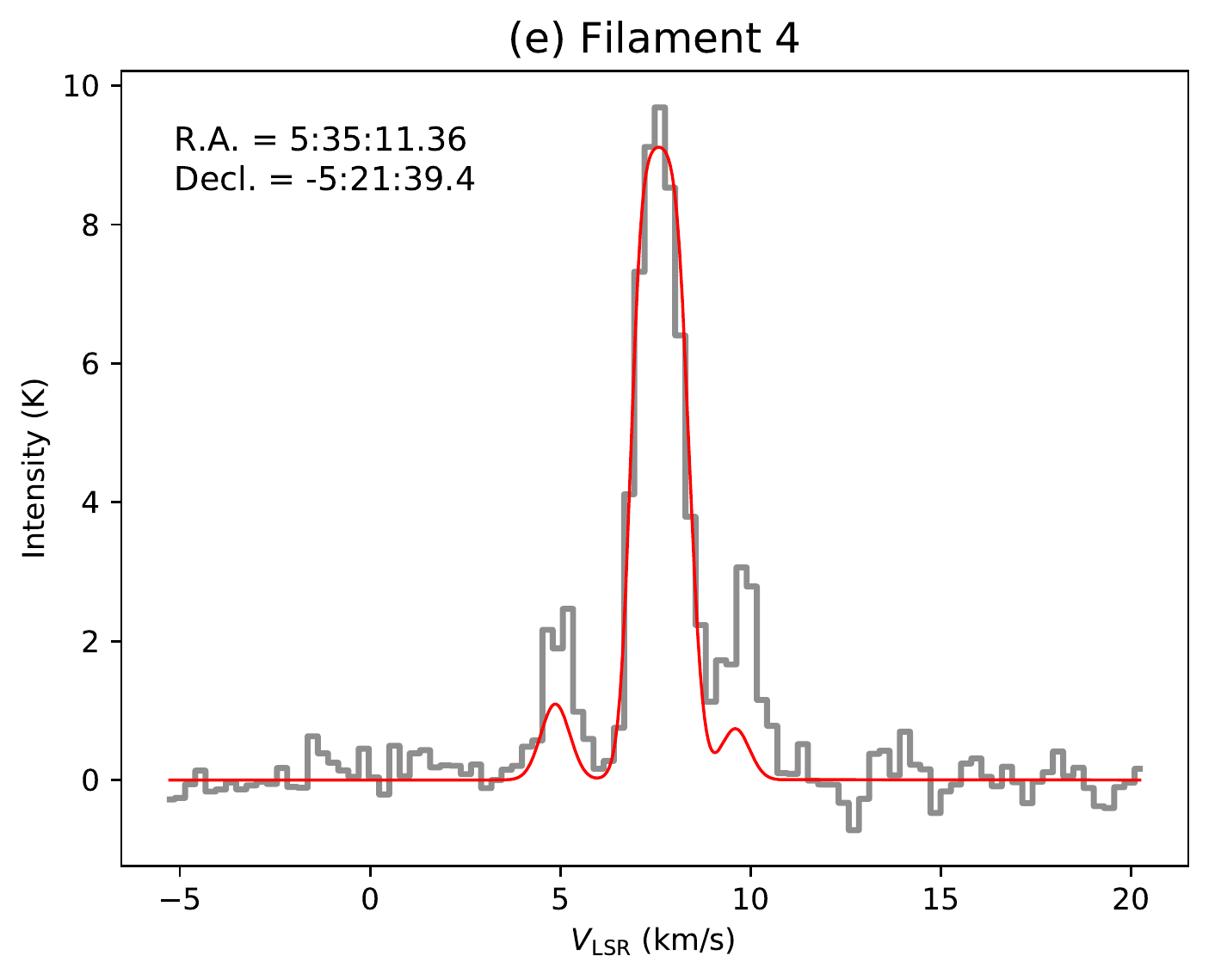} &\includegraphics[width=0.32\linewidth]{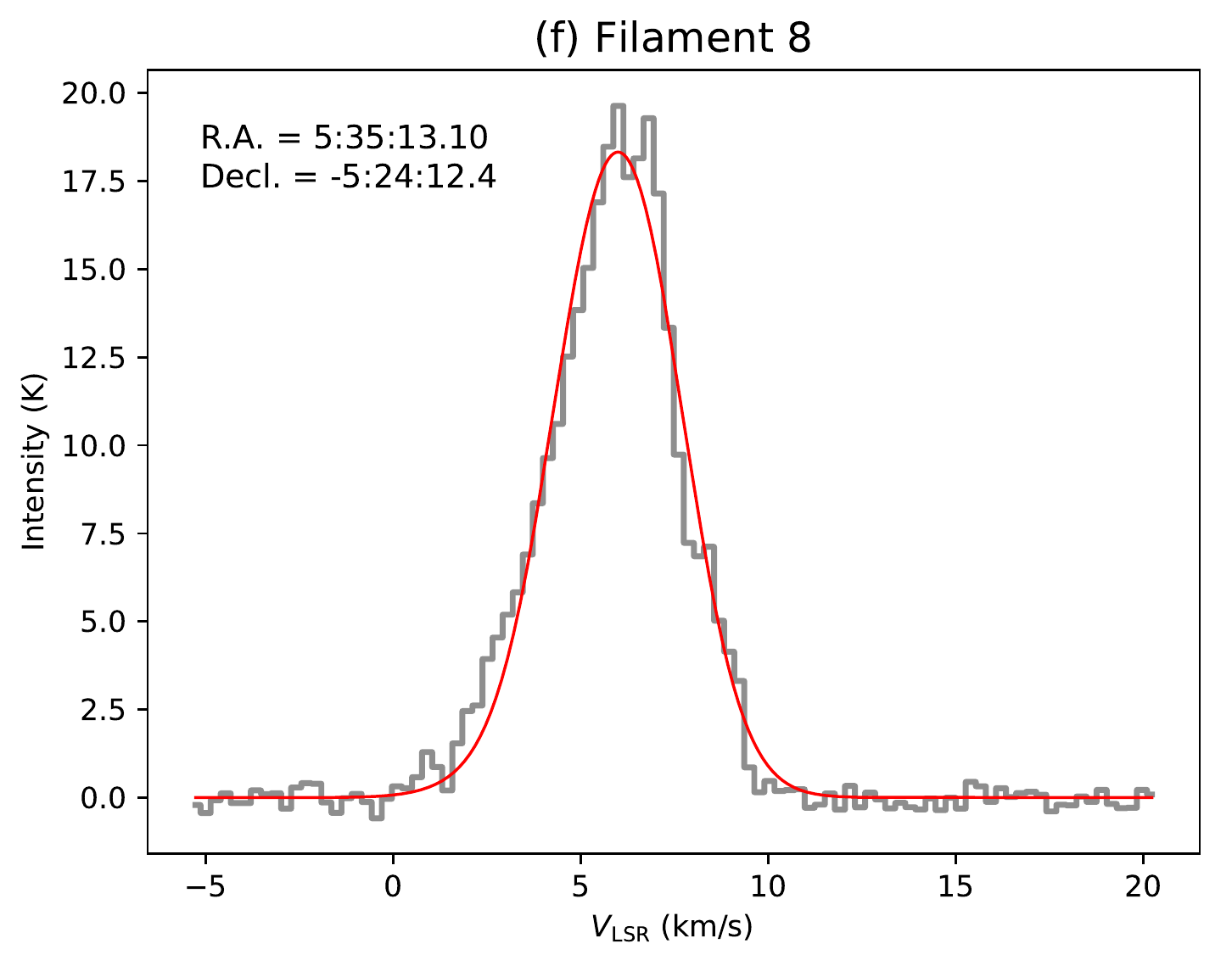}
\end{tabular}
\caption{Spectra of the six representative positions of the filaments marked in Figure \ref{fig:structure}a. Each spectrum (gray histogram) was obtained by averaging the $5''\times5''$ area. Red curves are the results of the hyperfine 
spectral fitting using the single-velocity component.  Part of the (a) main filament and (b) filament 6 have multiple velocity components along the line of sight. (c)--(f) Other filaments in OMC1 have single-component spectra. The deviation between the observed and fitted satellite components in (c)-—(e) suggests a non-LTE physical condition (see Section \ref{subsec::fitting}).}
\label{fig:spec}
\end{figure*}

Figure \ref{fig:spec} presents the observed spectra obtained by averaging the $5'' \times 5''$ regions marked in Figure \ref{fig:structure}a.
The best-fit single-velocity components (in red) are overlaid on the spectra. Note that the multiple velocity components are only limited to the regions near core 6, 7, 9 and the northern part of filament 6, as indicated by the dashed green boxes in Figure \ref{fig:structure}a. 
We applied the single-component fitting to the spectra of these regions, because two-component fitting requires too many parameters.
Inside the regions with multiple components, the fitted $V_{\rm LSR}$ are still dominated by the major components, although the fitted linewidths are highly affected by secondary components.

\begin{figure*}
\begin{minipage}{.48\linewidth}
\centering
\includegraphics[width=\linewidth]{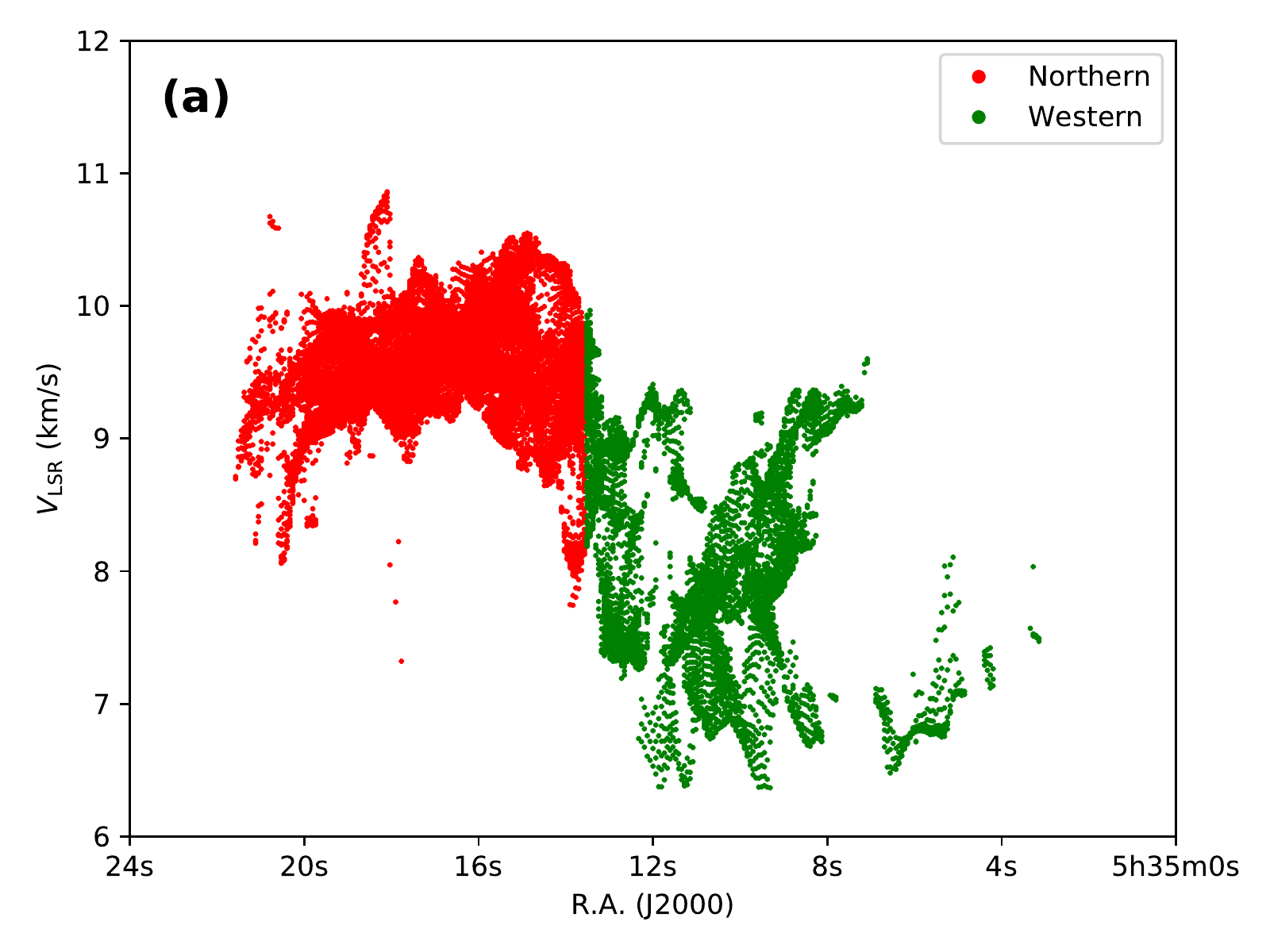}%\\
%(a)
\end{minipage}
\hfill
\begin{minipage}{.48\linewidth}
\centering
\includegraphics[width=\linewidth]{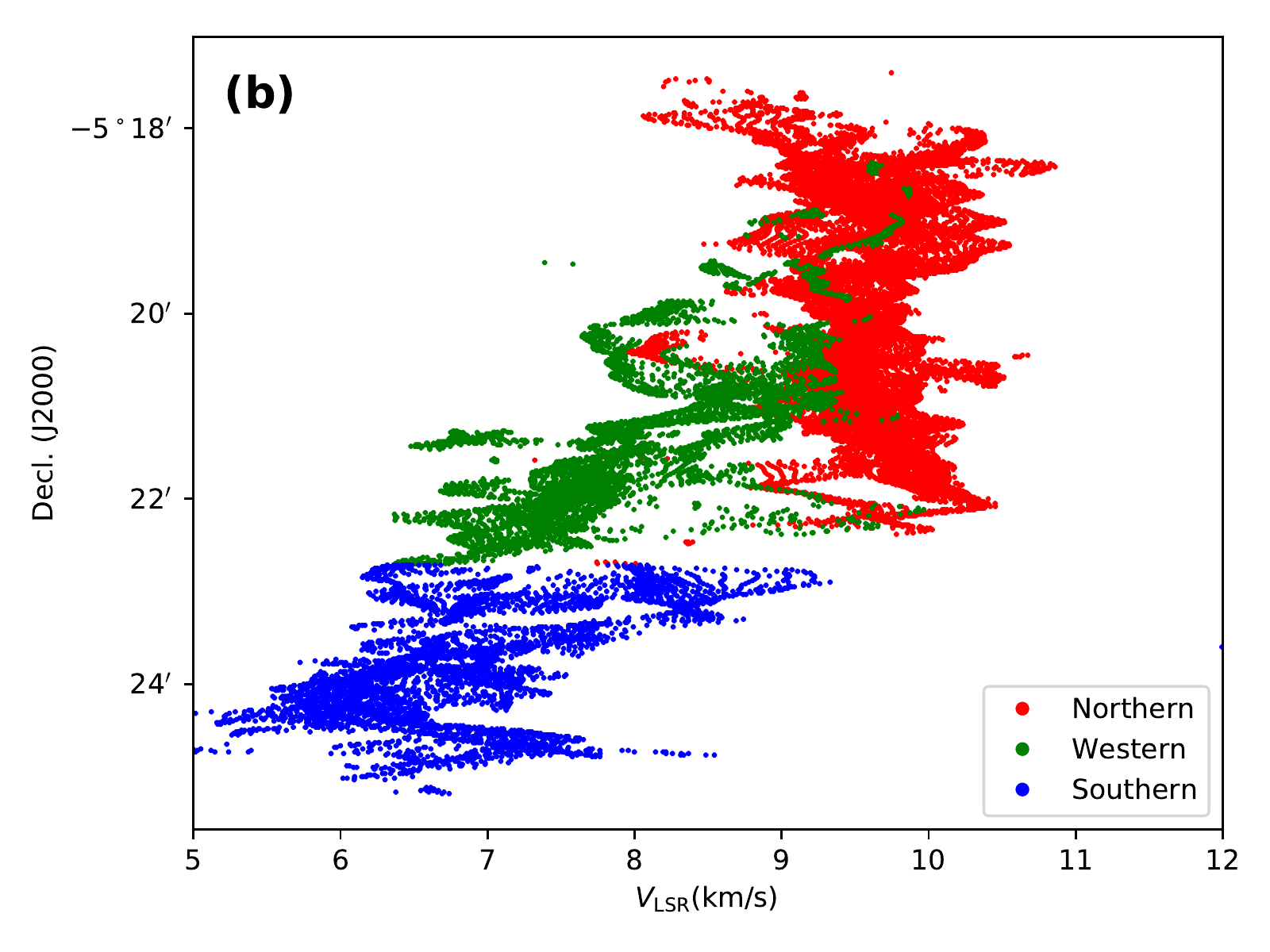}%\\
%(b)
\end{minipage}
\caption{$\rm N_2H^+$ 3--2 radial velocities of (a) the northern and western regions as a function of right ascension, and (b) all three regions as a function of declination.}
\label{fig:radec2vel}
\end{figure*}

Using the centroid velocities determined from 
the hyperfine fitting, we illustrate the velocity distribution along right ascension and declination, respectively.
Figure \ref{fig:radec2vel}a shows the radial velocities for all the $\rm{N_2H^+}$ 3--2 components fitted in the northern and western regions. The figure reveals a sharp velocity transition near R.A.(J2000) = $\rm 5^h 35^m 14^s$, where a clear boundary can also be seen in Figure \ref{fig:obs_combine}b between the northern and western regions.  Figure \ref{fig:radec2vel}b shows the radial velocities of all three regions as a function of declination. While the velocities of the northern region are almost constant, those of the western region decrease from north to south and continuously connected to those of the southern region. Such a velocity decrease from north to south in OMC1 has been reported in \citet{hacar_2017,omc1_alma}, and interpreted as the gravitational collapse of OMC1.
Figure \ref{fig:radec2vel}b shows that the gas in the western region is accelerated toward OMC1-South.

Comparing the fitting results with the observed spectra, we found that the observed satellite components are much brighter than those predicted by the LTE assumption (e.g. Figure \ref{fig:spec}c--e). In addition, when the line is optically thin ($\tau_{\rm tot} \ll 1$), the fitting cannot constrain $T_{\rm ex}$ and $\tau_{\rm tot}$, because Equation \ref{Tb_lte} will be reduced to
\begin{eqnarray}
&&T_{b}(v) \simeq \tau_{\rm tot} \cdot T_{\rm ex}
\label{eqn_Tmb_thin}
\end{eqnarray}
indicating that the $T_{\rm ex}$ and $\tau_{\rm tot}$ can be determined arbitrarily.
Therefore, we conduct non-LTE analysis in Section \ref{subsec::non-lte} to derive the physical conditions.

\begin{figure*}
\centering
\includegraphics[width=\linewidth]{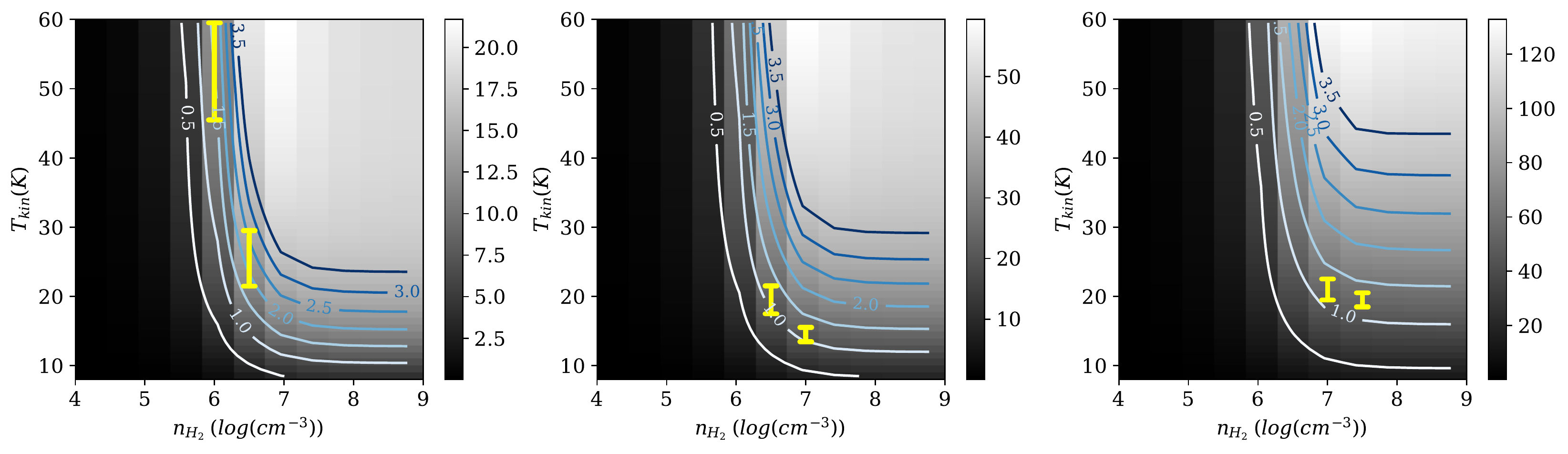}
\hfill
\caption{Slices in the non-LTE model grid for $N(\rm{N_2H^+})$/$\Delta v = 5\times10^{7}$, $5\times10^{7.5}$, and $5\times10^{8}$ $\rm s\ cm^{-3}$ from left to right, which corresponds to $N(\rm N_2H^+) = 10^{13}$, $3\times10^{13}$, and $10^{14}$ $\rm cm^{-2}$ for a region with $\Delta v = 2.0$ $\rm km\ s^{-1}$. The background in gray scale shows the integrated spectra model of $\rm N_2H^+$ 3--2, and the contour levels indicate the 3--2/1--0 intensity ratio model. The yellow bars from the left to right panels show possible solutions satisfying the conditions of the non-filament, low-intensity, and core regions, respectively (see Table \ref{tab:high-res}).}
\label{fig:non_lte}
\end{figure*}

\subsection{$N_2H^+$ RADEX Non-LTE Modeling} \label{subsec::non-lte}

Using RADEX \citep{radex}, a non-LTE radiative transfer code, we construct spectra models in $\rm N_2H^+$ 3--2 and 1--0. The synthetic spectra are constructed with the equation
\begin{eqnarray}
&&T_{b}(v) = \Psi \left( \frac{\sum J(T_{\rm ex}^i)\,\tau_i(v)}{\sum \tau_i(v)} - J(T_{\rm bg}) \right) \left( 1 - e^{-\sum \tau_i(v)} \right)
\nonumber \\
\label{Tb_non-lte}
\end{eqnarray}
where $T_{\rm ex}^i$ and $\tau_i(v)$ represent the excitation temperature and optical depth for all hyperfine components in the 3--2 or 1--0 transitions, and $\Psi$ is the beam filling factor. In our models, the beam filling factors are assumed to be unity. We also construct an intensity ratio model by dividing the integrated spectra model in $\rm N_2H^+$ 3--2 with that in $\rm N_2H^+$ 1--0. 

The constructed models can be represented by a three-dimensional grid, where the three axes are $\rm H_2$ density ($n_{\rm H_2}$) ranging from $10^4$ to $10^9$ $\rm cm^{-3}$, kinetic temperature ($T_{\rm kin}$) ranging from $8$ to $60$ K, and the ratio of $\rm N_2H^+$ column density to linewidth ($N(\rm{N_2H^+})$/$\Delta v$) ranging from $5\times10^{7}$ to $5\times10^{8}$ $\rm s\ cm^{-3}$. The step sizes of the grid are 1 K for $T_{\rm kin}$ and 0.5 in decimal log scale for both $N(\rm N_2H^+)$ and $n_{\rm H_2}$. 
As $N(\rm{N_2H^+})$/$\Delta v$ is the input parameter in RADEX, the estimation of $N(\rm N_2H^+)$ varies among regions with different linewidths. Based on our fitting results, the linewidth in OMC1 could vary from $\sim$ 1 to 3 $\rm km\ s^{-1}$.

Figure \ref{fig:non_lte} presents the constructed non-LTE model, where the gray-scale background shows the integrated spectra model in $\rm N_2H^+$ 3--2, and the contour levels show the 3--2/1--0 ratio model. By applying the observed integrated intensity and line ratio in this model, physical parameters ($n_{\rm H_2}$, $T_{\rm kin}$, and $N(\rm{N_2H^+})$/$\Delta v$) in different sub-regions can be constrained. For instance, the two bars on the first panel of Figure \ref{fig:non_lte} indicate the two possible solutions that satisfy
the conditions with an intensity of 7--15 $\rm K \cdot km\ s^{-1}$ and a line ratio of $2.2 \pm 0.4$. The corresponding physical conditions of the left and right bars are $n_{\rm H_2} = 10^6$ $\rm cm^{-3}$ and $T_{\rm kin}$ = 45--60 K, and $n_{\rm H_2} = 3 \times 10^6$ $\rm cm^{-3}$ and $T_{\rm kin}$ = 21--30 K, respectively.

\begin{table*}
\centering
\caption{Parameters for non-LTE analysis \label{tab:high-res}}
\begin{tabular}{lccc}
  \hline\hline
  &Core Regions &Low Intensity &Non-filament\\
  &($>50$ $\rm K \cdot km\ s^{-1}$) &Regions &Regions \\
  \hline
  $n_{\rm H_2}$ ($\rm cm^{-3}$) &$10^7$ or $3 \times 10^7$ &$3 \times 10^6$ or $10^7$ &$10^6$ or $3 \times 10^6$ \\
  $T_{\rm kin}$ (K) &$19$--$23$ or $18$--$20$ &$17$--$22$ or $13$--$16$ &$>45$ or $21$--$30$ \\
  $N(N_2H^+)/\Delta v$ ($\rm s\ cm^{-3}$) &$5\times10^{8}$ &$1.5 \times 10^{8}$ &$5\times10^{7}$ \\
 Intensity ($\rm K \cdot km\ s^{-1}$) &50--60 &20--40 &7--15 \\
  Typical Ratio &$1 \pm 0.3$ &$1 \pm 0.3$ &$2.2 \pm 0.4$ \\
  \hline
\end{tabular}
\end{table*}

We use our SMA + SMT data in $\rm N_2H^+$ 3--2 together with the ALMA + IRAM 30-m data in $\rm N_2H^+$ 1--0 for the non-LTE analysis, where the scale size is $5.4''$ ($\sim0.01$ pc).
Based on the 3--2 moment 0 map, we define three regions for analysis. The first region is defined as the core regions with integrated intensity 50--60 $\rm K \cdot km\ s^{-1}$ and a 3--2/1--0 ratio of $1 \pm 0.3$. The second region is the lower intensity regions inside the filaments (20--40 $\rm K \cdot km\ s^{-1}$) with a similar line ratio of $1 \pm 0.3$, and the third region is the non-filament region with low intensities (7--15 $\rm K \cdot km\ s^{-1}$) and a higher line ratio of $2.2 \pm 0.4$.
The spectra averaged in these sub-regions are compared with the model spectra. The derived physical parameters for the non-LTE analysis together with the criteria of the three sub-regions are listed in Table \ref{tab:high-res}.

By comparing between the filament and non-filament regions, we find that the filament regions have a higher density of $\sim 10^7$ $\rm cm^{-3}$ and a lower temperature of $\sim 15$--$20$ K than the non-filament regions. As the major heating sources may come from outside the filaments, it is likely that the dense gas in the filaments could block the outer radiation, leading to a lower temperature in the filaments (see Section \ref{subsec:heating} for further discussion). 
Inside the filaments, the core regions have a higher $\rm N_2H^+$ column density than the low intensity regions, if we assume similar linewidths in these regions. Also, the volume density of the core regions are generally higher than the low intensity regions. 
On the other hand, there is no significant difference in temperature between the core and the low intensity regions, although the derived $T_{\rm kin}$ is slightly higher in the core regions.

Using the volume densities determined from the non-LTE analysis and the filament widths, we have estimated the masses of the cores and the line masses of the filaments under the assumption of uniform cylindrical filaments. 
We adopted this method instead of using the column density of N$_2$H$^+$ because the fractional abundance of N$_2$H$^+$ is highly uncertain in the regions close to Orion KL and because the inclination of each filament is also uncertain.
The density is assumed to be $n_{\rm H_2} = 3\times10^6$ $\rm cm^{-3}$ for the filaments without cores. For those with cores, i.e. main filament, east filament, and filament 8, we use two possible core densities (i.e. $10^7$ or $\rm 3\times10^7 cm^{-3}$) determined in this section for the core regions and $n_{\rm H_2} = 3\times10^6$ $\rm cm^{-3}$ for the regions outside the cores.
If we adopt the higher density solution, i.e. $10^7$ $\rm cm^{-3}$, for the low intensity regions of the filaments, the line masses of the filaments could be higher by a factor of $\sim$3. 
On the other hand, if we adopt the radial density profile of the isothermal cylinder with the same central density \citep[e.g.][]{Mcrit_2}, the line mass included in the radius of 0.01--0.015 pc (i.e. half of the filament width) becomes lower by a factor of two.
Properties of the identified filaments and cores are summarized in Table \ref{tab:fil_ident} and Table \ref{tab:core_ident}, respectively, and will be discussed in Section \ref{subsec:core_formation}.

\subsection{Gas Kinematics of the Filaments} \label{subsec:kinematics}
Characterizing the gas motion inside the filaments requires the studies of velocity structure. We investigate the radial velocity fields along both the major and minor axes of the filaments in OMC1, and compare our results with existing filament formation model and core formation model. Since the main filament and the east filament are identified with core fragmentation, we focus on the analyses of these two filaments.

\begin{figure*}
\begin{minipage}{.48\linewidth}
\centering
\includegraphics[width=\linewidth]{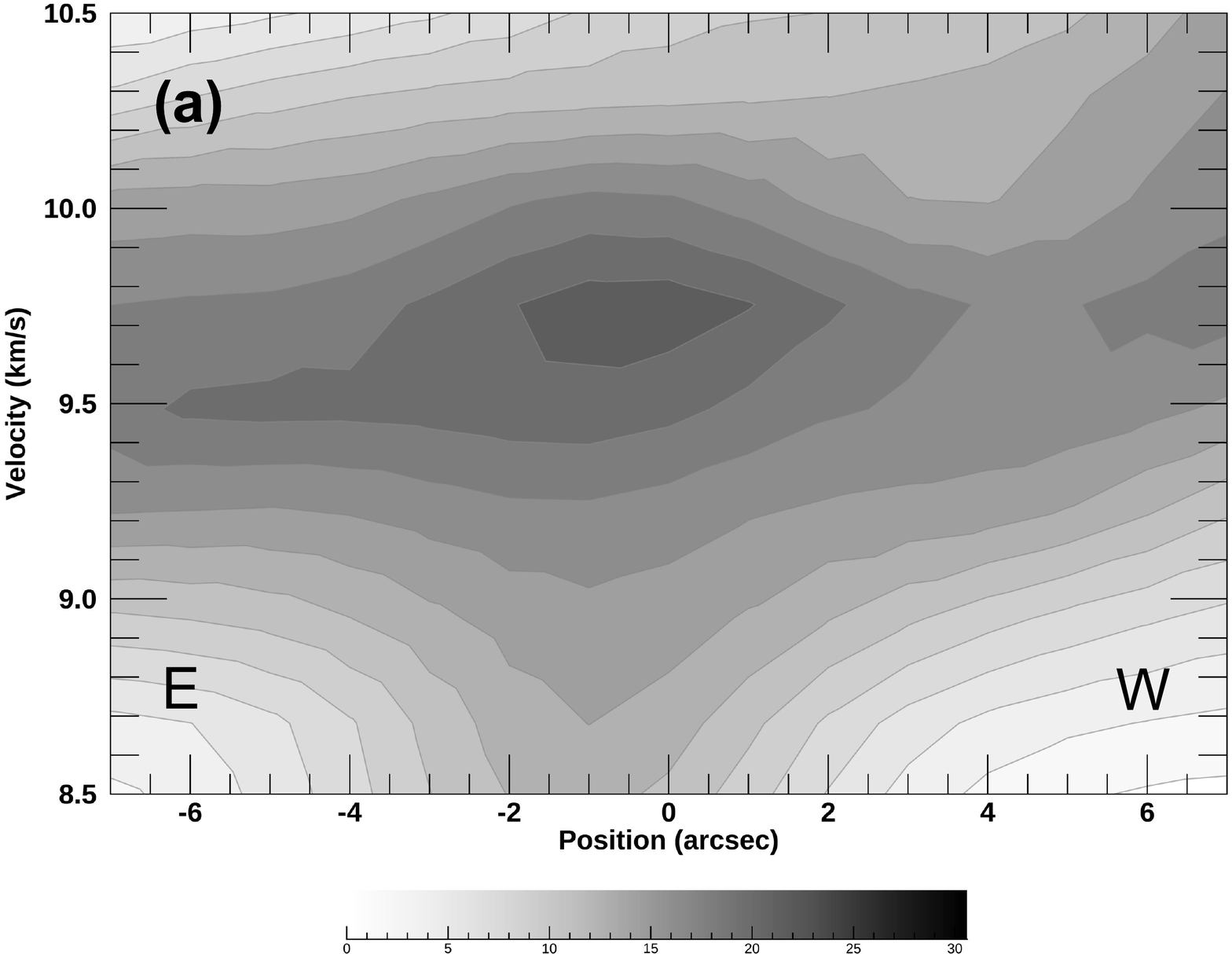}%\\
%(a)
\end{minipage}
\hfill
\begin{minipage}{.48\linewidth}
\centering
\includegraphics[width=\linewidth]{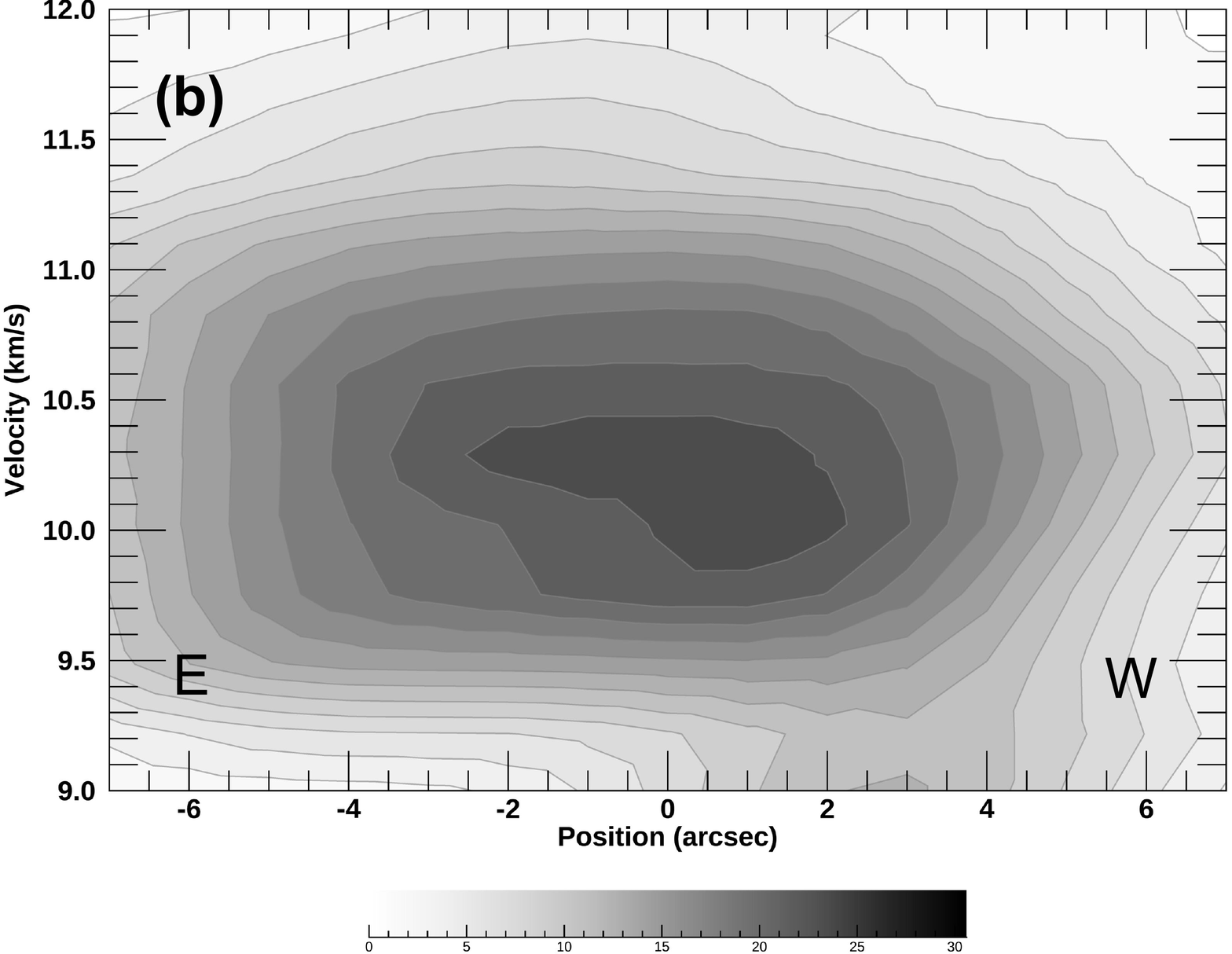}%\\
%(b)
\end{minipage}
\caption{Position--Velocity (P--V) diagrams across (a) core 8 and (b) core 4 inside the east filament.}
\label{fig:v_minor}
\end{figure*}

\subsubsection{Minor-Axis Analysis} \label{subsubsec::minor}

Systematic velocity gradients perpendicular to the filaments have been observed in the filaments of both low- and high-mass star-forming regions \citep[e.g.][]{serpens_ngc1333,schneider_2010,beuther_2015}. Such velocity gradients across the filaments can be explained by the projection of gas accretion toward the filament axes \citep{serpens_ngc1333} or the rotation of filaments \citep{fil_rotate}. In order to assess whether the OMC1 filaments show similar features, we analyze the velocity fields along the minor axes.

In the main and the east filaments, there is no systematic velocity gradient along the minor axes. However, local velocity gradients of $\sim 0.3$ $\rm km\ s^{-1}$ are observed in core 4 and 8 in the east filament. On the other hand, there is no significant velocity gradient in core 5 and 10.
The velocity gradients in core 6, 7, and 9 in the northern part of the main filament are unclear due to the secondary velocity component along the line of sight.
Figure \ref{fig:v_minor} shows Position--Velocity (P--V) diagrams across core 8 and core 4. 
The directions of velocity gradients across these cores are not consistent; the velocity increases from east to west in core 8, while it decreases in core 4. Such velocity gradients with different directions along the same filament have also been observed along the massive DR21 filament, although its origin is still unclear \citep{schneider_2010}. 

Using the effective radii determined in Section \ref{subsubsec::core_ident}, i.e. $R_{\rm eff} = 5.1''$ and $3.6''$ for core 4 and core 8, respectively, the velocities at $R_{\rm eff}$ were determined from the spectral fitting. 
Then, the velocity gradient ($\nabla V$) along the east-western direction was estimated using the velocity difference at the core boundaries and the effective diameter. 
Assuming a linear velocity variation, core 8 and core 4 have $\nabla V = 17.2$ $\rm km\ s^{-1}\ pc^{-1}$ and $\nabla V = -11.3$ $\rm km\ s^{-1}\ pc^{-1}$, respectively. The origin of the observed velocity gradients with different directions are likely to be the local effects such as rotating motions or unresolved multiple components. If the observed velocity fields come from rotations, the rotational energy $E_{\rm rot}$ can be estimated under the assumption of solid-body rotation as \citep{rotation} 
\begin{eqnarray}
&&E_{\rm rot} = \frac{1}{2} I \Omega^2 = \frac{1}{2} \left[\frac{2}{3} MR^2 \left( \frac{3-\alpha}{5-\alpha} \right)\right] \left(\frac{|\nabla V|}{\sin{i}}\right)^2
\label{rot_energy}
\end{eqnarray}
where $i$ is the inclination angle of the rotational axis to the line of sight, and $\alpha$ indicates a power-law density profile of $\rho \propto r^{-\alpha}$.
The gravitational energy can also be derived as
\begin{eqnarray}
&&E_{\rm grav} = -\frac{GM^2}{R} \left( \frac{3-\alpha}{5-2\alpha} \right)
\label{grav_energy}
\end{eqnarray}

To estimate the ratio of rotational to gravitational energy ($\beta_{\rm rot}$) for these cores, we assume a uniform density profile, i.e. $\alpha = 0$, and a random-averaged inclination angle of $\langle \sin i \rangle = \pi / 4$. It turns out that core 8 has $\beta_{\rm rot}$ ranging from 0.04 to 0.11, and core 4 has $\beta_{\rm rot}$ ranging from 0.02 to 0.06. This means that rotations of these cores are significant but not dominant in the energetics.

\subsubsection{Major-Axis Analysis} \label{subsubsec::major}

\begin{figure*}
\begin{minipage}{.5\linewidth}
\centering
\includegraphics[width=\linewidth]{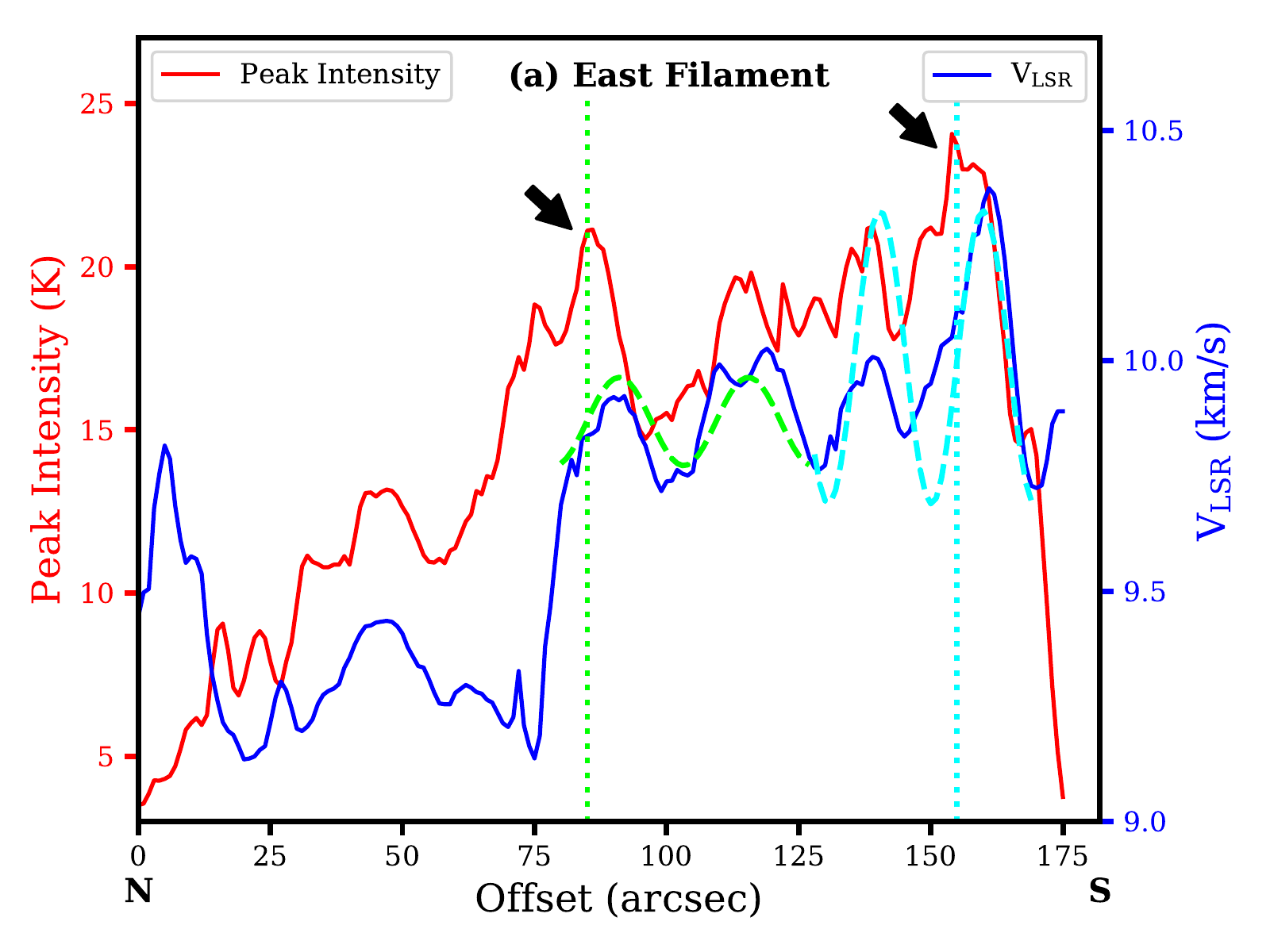}%\\
%(a)
\end{minipage}
\hfill
\begin{minipage}{.5\linewidth}
\centering
\includegraphics[width=\linewidth]{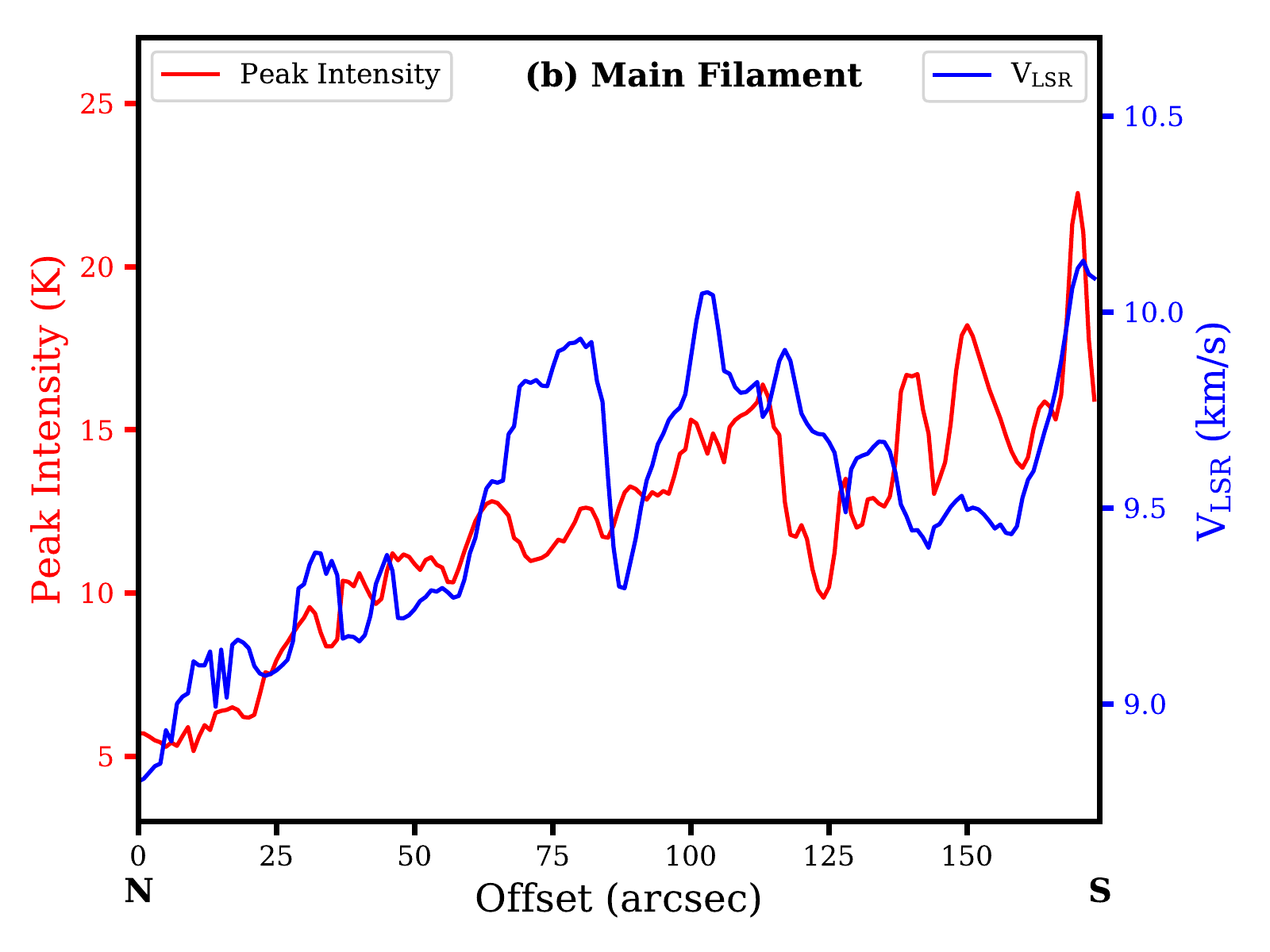}%\\
%(b)
\end{minipage}
\caption{Comparison between the intensity and velocity variations along the major axis of the (a) east filament and the (b) main filament. In the left panel, the arrows indicate two intensity peaks which correspond to core 8 and 4, showing positional shifts relative to their velocity peaks; the dashed curves are sinusoidal fits to the velocity variation, and the two vertical lines show the locations shifted by $\lambda/4$ from the fitted sinusoidal peaks of core 8 and 4. Offset 0 is at the northernmost position of the filament, i.e. (R.A., Decl.) = ($\rm 5^h 35^m 19^s.4$,$-5^\circ 19' 29''.43$) for the east filament and ($\rm 5^h 35^m 20^s.4$,$-5^\circ 17' 50''.43$) for the main filament.}
\label{fig:v_major}
\end{figure*}

The velocity fields along the major axes of the filaments are also analyzed. Using the central velocity ($V_{\rm LSR}$) and the peak intensity determined from the hyperfine spectral fitting (see Section \ref{subsec::fitting}), we plot the intensity and velocity variations along the filament major axis in Figure \ref{fig:v_major}, where the horizontal axes represent the offset (from north to south) in arcseconds, and the vertical axes show the peak intensity on the left and the $V_{\rm LSR}$ on the right.

Analysis on the east filament (see Figure \ref{fig:v_major}a) reveals the oscillations in both intensity and velocity.
It is found that there are positional shifts between the intensity and velocity peaks toward core 8 and 4. This is similar to the feature observed in two of the filaments in L1517 \citep{Hacar_Tafalla_2011}. The velocity oscillation along the filament can be related to the core-forming motions. If the gas flow is converging to the center of the core, its velocity with respect to the one at the core center is positive in one side and negative in the other side of the density peak. According to the kinematic model proposed in \citet{Hacar_Tafalla_2011}, where sinusoidal perturbations were assumed for both density and velocity, a $\lambda / 4$ phase shift between the two distributions is predicted. In Figure \ref{fig:v_major}a, we show the sinusoidal fits to the velocity oscillation in the east filament. In addition, the locations of the intensity peaks toward core 8 and 4 match well with the $\lambda / 4$ shift from the two corresponding sinusoidal peaks.

%On the other hand, relations between the intensity and velocity variations are unclear in the main filament (see Figure \ref{fig:v_major}b). 
Figure \ref{fig:v_major}b shows the intensity and velocity plot for the main filament.
Although there is a secondary velocity component toward part of this filament, the velocities determined from the fitting represent the ones of the major component.
In the main filament, the relations between the intensity and velocity variations are unclear.
This is probably because of the evolutionary stage of the main filament. Previous 1.3 mm observations with the SMA \citep{sma_north_fil} revealed CO molecular outflows associated with some cores in the main filament, including core 5. It is therefore likely that some of the cores in the main filament have already harbor young protostars in the Class 0 stage.

In contrast, no evidence of protostars have been observed in the east filament, and the positional shift between its intensity and velocity peaks may indicate that core formation is still ongoing in the east filament. As the evolutionary stage of different filaments can vary \citep{sf_fil_model}, it is possible that the east filament is in an earlier evolutionary phase than the main filament with star formation signature.

\section{Discussion} \label{sec:discussion}

\subsection{Filament and Core Properties} \label{subsec:core_formation}

%Using the volume densities determined from the non-LTE analysis and the filament widths, the line masses of the filaments have been derived under the assumption of uniform cylindrical filaments. The density is assumed to be $n_{\rm H_2} = 3\times10^6$ $\rm cm^{-3}$ for the filaments without cores. For those with cores, i.e. main filament, east filament, and filament 8, we use two possible core densities (i.e. $10^7$ or $\rm 3\times10^7 cm^{-3}$) determined in Section \ref{subsubsec::high-res} for the core regions and $n_{\rm H_2} = 3\times10^6$ $\rm cm^{-3}$ for the regions outside the cores.

As shown in Table \ref{tab:fil_ident}, all three filaments with core fragmentation have line masses $\gtrsim 80$ $\rm M_\odot\ pc^{-1}$. 
In contrast, the filaments with lower line masses such as filament 3, 4, and 9 do not contain cores.
However, three filaments having the line masses close to 80 $\rm M_\odot\ pc^{-1}$ (filament 1, 2, and 5) and filament 6 with the largest line mass do not contain any cores. Therefore, although line masses are often used as an indicator of the star formation stage of a filament \citep{Heitsch_2013,Palmeirim_2013,Li_2014}, it may not be a conclusive discriminator, which is also stated in \citet{serpens_ngc1333}.

To investigate the internal dynamics of the filaments, we derive the non-thermal velocity dispersion ($\Delta v_{\rm nt}$) by using the equation
\begin{eqnarray}
&&\Delta v_{\rm nt} = \sqrt{\frac{(\Delta v)^2}{8\ln{2}} - \frac{k \cdot T_{\rm kin}}{\mu(\rm{N_2H^+})}}
\label{def_v_nt}
\end{eqnarray}
where $\Delta v$ is the linewidth in FWHM obtained from the hyperfine fitting, and $\mu$ is the molecular mass.
The non-thermal velocity dispersion can be compared with the thermal sound speed
\begin{eqnarray}
&&c_s(T) = \sqrt{\frac{k \cdot T_{\rm kin}}{\mu(\rm{H_2})}}
\label{def_v_th}
\end{eqnarray}
Then, the critical line mass for an infinite filament in hydrostatic equilibrium can be calculated as \citep{Mcrit_1,Mcrit_2}
\begin{eqnarray}
&&M_{\rm crit} = \frac{2(\Delta v_{\rm eff})^2}{G} = \frac{2}{G} \left[c_s(T)^2 + (\Delta v_{\rm nt})^2 \right]
\label{def_Mcrit}
\end{eqnarray}
where $\Delta v_{\rm eff}$ is defined as the effective velocity dispersion considering both thermal and non-thermal effects.

Based on the non-LTE results in Section \ref{subsec::non-lte}, we take $T_{\rm kin} = 20$ K for estimation, which leads to $c_s(T) = 0.287$ $\rm km\ s^{-1}$. 
In the case of the purely thermally-supported filament, $\rm M_{crit} = 38.4\ M_\odot\ pc^{-1}$. 
Due to the rather large non-thermal velocity dispersion, the critical line masses listed in the OMC1 filaments are $\sim$2--4 times larger than the purely thermally-supported case.
Table \ref{tab:fil_ident} reveals that 7 of the 11 filaments have $\rm 0.5 \le M_{lin}/M_{crit} \le 1.5$, suggesting that most of the filaments are gravitationally bound. 
Filament 3, 4, and 9 having low line masses are found to have $\rm M_{lin}/M_{crit} < 0.5$ and thus may be gravitationally unbound. 
On the other hand, filament 8 has $\rm M_{lin}/M_{crit} < 0.5$ even though this filament contains cores. This filament resides in OMC1-South, where a cluster of young stellar objects (YSOs) have already been formed \citep{zapata_2004,zapata_2006}. Due to the powerful outflows from these YSOs, the gas in this region is highly turbulent. Thus, filament 8 has the largest linewidth of $\sim 3$ $\rm km\ s^{-1}$ among all the filaments, leading to a high $\rm M_{crit}$ and a low $\rm M_{lin}/M_{crit}$ ratio. The low $\rm M_{lin}/M_{crit}$ value could imply that filament 8 is in the phase of disruption. Another possibility is the high external pressure in this region. If the filament is 
confined by external pressure, the filament is prone to fragmentation even though its line mass is smaller than the critical line mass \citep[e.g.][]{2012A&A...542A..77F}. Since filament 8 in OMC1-South is adjacent to the Orion Nebula Cluster, the high external pressure from hot and diffuse gas in the cluster could lead the core 
formation in this filament. 

Our results show that the filaments in OMC1 have $\rm M_{lin}/M_{crit} < 1.5$, which is consistent with that of \citet{omc1_alma} based on their $\rm N_2H^+$ 1--0 data. However, even though the derived $\rm M_{lin}/M_{crit}$ ratios are similar, both the $\rm M_{lin}$ and $\rm M_{crit}$ determined from our data are higher than those of \citet{omc1_alma} by a factor of few.
While they used $\rm N_2H^+$ intensities to estimate column densities of $\rm H_2$, {we use the volume densities of $\rm H_2$ derived from the non-LTE analysis.
The empirical relation between N$_2$H$^+$ intensities to H$_2$ column densities used by \citet{omc1_alma} has large scatter.
The inclination of each filament is also uncertain.}
On the other hand, the uncertainty in inclination does not affect our estimation.
However, the $\rm M_{lin}$ derived under our assumption of uniform density inside the filaments (excluding the cores) might be overestimated if the filaments have radial density profiles. For the difference in $\rm M_{crit}$, it is likely resulted from the difference between the linewidths observed in $\rm N_2H^+$ 3--2  ($\sim$1.0 km s$^{-1}$) and 1--0.
This is partly because of the difference in the beamsize of our  measurement ($5.4''$) and \citet{omc1_alma} ($4.5''$); the spectra observed with the larger beam contain the nonthermal motion in the larger area.
The hyperfine components of 3--2, which is much more complicated than those of 1--0, introduce additional uncertainty in the derived linewidth if the relative intensities of the hyperfine components are different from the LTE values.

Table \ref{tab:core_ident} shows that most of the cores have masses ($\rm M_{core}$) ranging from $1$--$10$ $\rm M_\odot$.
Both the sizes and the masses of these cores are larger than those determined by \citet{sma_north_fil}, because multiple smaller-scale cores were revealed in their 1.3 mm continuum data using the SMA.
These masses can also be compared with the masses inferred by the virial theorem. 
By assuming a uniform density in the cores, the virial mass ($\rm M_{vir}$) can be derived as 
\begin{eqnarray}
&&M_{\rm vir} = \frac{5\ R_{\rm eff} (\Delta v)^2}{8\ G \ln{2}}
\label{def_Mvir}
\end{eqnarray}
where $R_{\rm eff}$ is an effective radius of the core, and $\Delta v$ is a typical linewidth.
The linewidth $\Delta v$ was determined from the hyperfine fitting except for core 6, 7, and 9 having multiple velocity components along the line of sight.

As shown in Table \ref{tab:core_ident}, the measured core masses are similar to the calculated virial masses, indicating that most of the cores are gravitationally bound. 
One of the cores, core 5, already shows the signatures of star formation; toward this core, there is an infrared source and a clear bipolar outflow in CO \citep{sma_north_fil}.
The velocity dispersion as well as $\rm M_{vir}$ are large in the southern cores i.e. core 1 and 2, because of the strong turbulence from the stellar activities in OMC1-South \citep{zapata_2006,2018ApJ...855...24P}.

\subsection{External Heating from High-mass Stars} \label{subsec:heating}

The $\rm N_2H^+$ 3--2/1--0 ratio map presented in Figure \ref{fig:ratio} shows higher ratios in the eastern part of OMC1. From the non-LTE model shown in Figure \ref{fig:non_lte}, regions with a higher intensity ratio generally indicates a higher kinetic temperature. This suggests that the eastern OMC1 has higher temperatures comparing with the remaining area.

We find that the overall distribution of the high-ratio gas is similar to that of the CN and $\rm C_2H$ molecules presented in \citet{ungerechts_1997} and \citet{melnick_2011}. Since CN and $\rm C_2H$ are sensitive to the presence of UV raidation \citep{fuente_1993,stauber_2004}, it is likely that the higher temperatures in the eastern OMC1 are caused by the UV heating from the high-mass stars in M42. For example, UV photons from the $\rm \theta^1$ Ori C at (R.A., Decl.) = ($\rm 5^h 35^m 16^s.5$,$-5^\circ 23' 22''.8$), which is southeast to the Orion KL, could be one of the major heating sources.
In addition, the [C{\small \ I}]/CO intensity ratio map in \citet{orion_ci_mapping} shows a ratio peak of $\sim 0.17$ around the position (R.A., Decl.) = ($\rm 5^h 35^m 20^s$,$-5^\circ 18' 30''$), implying that UV radiation also contributes to the heating of this region. Possible heating sources include the exciting star of M43---NU Ori at (R.A., Decl.) = ($\rm 5^h 35^m 31^s.0$, $-5^\circ 16' 12''$), which is northeast to the OMC1 region.

The external heating scenario also explains the 3--2/1--0 ratios inside and outside the filaments. As shown in Figure \ref{fig:ratio}, heating features are seen only outside the filament regions, while temperatures inside the filaments remain significantly lower.
\citet{omc1_nh3} reported the variation of NH$_3$ $(J, K)$ = (2, 2) to (1, 1) line intensity ratio in the filaments; the higher ratio (i.e. higher temperature) gas appears between the emission peaks with lower ratio (i.e. lower temperature).
Such a patchy heating pattern in the filaments does not appear in the N$_2$H$^+$.
There is no significant difference in the $\rm N_2H^+$ 3--2/1--0 ratio between the core regions and the low intensity regions in the filaments.
This is probably because the N$_2$H$^+$ lines having higher critical densities than the NH$_3$ trace the inner and denser part of the filaments where the gas is shielded well from the external radiation. 

Apart from external UV heating, local heating from Orion KL has been discussed on the basis of previous observations \citep{tang_2018,bally_alma_co,zapata_2011,omc1_nh3}. However, since $\rm N_2H^+$ emission is missing toward Orion KL, local heating around the KL core ($\sim 100$ K) is not observed in our data. The 3--2/1--0 ratio at the position of core 4, $\sim 2.0$, is significantly higher than other regions of the east filament. This could be the effect of the local heating from Orion KL.

\subsection{Global Collapse of OMC1} \label{subsec:collapse}

Recent studies have shown that large-scale dynamical collapse are important in high-mass star-forming regions \citep[e.g.][]{hartmann_2007,hacar_2017}.
The snapshots from the magneto-hydrodynamics (MHD) simulation of globally collapsing clouds presented in \citet{schneider_2010} and \citet{peretto_2013} reveal multiple filaments and sharp boundaries of radial velocity changes that separate the cloud into several regions, and these regions and filaments converge toward the massive core at the center. 
The radial velocity distribution of OMC1 shows the trimodal pattern centered near Orion KL with the sharp boundaries between the northern and western regions (Figure  \ref{fig:radec2vel}a) and the northern and southern regions (Figure  \ref{fig:radec2vel}b). In addition, the radial velocities in the western region monotonically decrease from north to south, and continue to the velocity gradient in the southern region (Figure \ref{fig:radec2vel}b).
This velocity gradient corresponds to the part of the V-shaped velocity structure centered around the OMC1-South, which is interpreted as the presence of accelerated gas motion inflowing toward the Orion Nebula Cluster \citep{hacar_2017}.
Our results have revealed that the gas in the western and southern regions contributes to this inflow.

Such a global collapse picture is also supported by the morphology of the magnetic field in OMC1.
The magnetic field in OMC1 revealed by the B-Fields In Star-Forming Region Observations (BISTRO) survey using the James Clerk Maxwell Telescope (JCMT) shows a well-ordered U-shaped structure, which can be explained by the distortion of an initially cylindrically-symmetric magnetic field due to large-scale gravitational collapse \citep{jcmt_bistro}.
Interestingly, the filaments in the northern and southern regions are almost perpendicular to the local magnetic field direction, while those in the western region such as filament 3, 4, and 5 are aligned along the magnetic field.
This implies that the filaments in the western region are feeding material along the magnetic field lines to the Orion Nebula Cluster, as predicted from the MHD simulation of global collapse \citep{schneider_2010}.

\section{Conclusion} \label{sec:conclusion}

We conducted the $\rm N_2H^+$ 3--2 line observations toward the OMC1 region using the SMA and SMT. The SMA data are combined with the SMT data in order to recover the spatially extended emission. The filaments and cores in OMC1 have been identified, and their physical properties are derived. Using the $\rm N_2H^+$ 1--0 data provided by \citet{omc1_alma}, we conducted the non-LTE analysis, and determined the kinetic temperatures and $\rm H_2$ densities of the filaments and cores. 
We also examine the gas kinematics inside the two prominent (main and east) filaments and compare them with the filament/core formation models. The main results are summarized as the following:

\begin{enumerate}
\item The combined SMA and SMT image in $\rm N_2H^+$ 3--2 reveals multiple filamentary structure having typical widths of 0.02--0.03 pc. In total , 11 filaments and 10 cores are identified. 
Cores are found in the three filaments with the line masses $\gtrsim 80$ $\rm M_\odot\ pc^{-1}$. 
The masses of the cores are in the range of $1$--$10$ $\rm{M_\odot}$ except the most massive one with $>$14 $M_{\odot}$ in OMC1-South. Up to $\sim 65\%$ of the filaments are gravitationally bound, which could be current or future sites for star formation.

\item The result of non-LTE analysis shows that the kinetic temperature is enhanced in the eastern part of OMC1. This is probably because of the external heating from high-mass stars in M42 and M43 (e.g. $\theta^1$ Ori C and NU Ori). 
It is found that the filament regions with higher densities of $n_{\rm H_2}\sim 10^7$ $\rm cm^{-3}$ have lower temperatures ($T_{\rm kin}\sim$ 15--20 K) than their surrounding regions. The lower temperatures in the filaments can be explained by the shielding from the external heating by dense gas.

\item 
The moment 1 image reveals that OMC1 consists of three sub-regions with different radial velocities divided by the sharp velocity transitions.
Three sub-regions intersect with each other at the location of Orion KL.
The radial velocities in the western region monotonically decreases from north to south, and continue to that in the southern region.
The observed velocity structure suggests the presence of global gas flow toward the Orion Nebula Cluster.
Such a global collapse is also supported by the observed morphology of magnetic fields and large scale gas kinematics in the integral-shaped filament.

\item 
Non-thermal motion plays important roles in the OMC1 filaments. 
There is no systematic velocity gradient along the minor axes of the OMC1 filaments.
Although the velocity gradient of $\sim 0.3$ $\rm km\ s^{-1}$ is observed in the east filament, the direction of the velocity gradient is different at different locations.

\item 
Two cores in the east filament show positional shifts between their intensity and velocity peaks along the filament major axis. It is possible that core formation is still ongoing in this filament. 
On the other hand, there is no such positional shift in the main filament.
It is therefore likely that the east filament is in an earlier evolutionary phase than the main filament, which shows the signatures of star formation.
\end{enumerate}

\acknowledgments
We thank the staffs of the SMA and the SMT for the operation of our observations and their support in data reduction. We thank Dr. T.-H. Hsieh for helping us with the SMT observations. The SMA is a joint project between the Smithsonian Astrophysical Observatory and the Academia Sinica Institute of Astronomy and Astrophysics and is funded by the Smithsonian Institution and the Academia Sinica. N.~H. acknowledges a grant from the Ministry of Science and Technology (MoST) of Taiwan (MoST 108-2112-M-001-017).   

\vspace{5mm}
\facilities{SMA, SMT}
\software{MIR/IDL (\url{https://www.cfa.harvard.edu/rtdc/SMAdata/process/mir/}), MIRIAD \citep{miriad}, GILDAS/CLASS (\url{http://www.iram.fr/IRAMFR/GILDAS}), 
FilFinder \citep{filfinder}, 
Clumpfind \citep{clfind}
}

% \appendix

\end{document}